\documentclass[aps,prd,twocolumn,preprintnumbers,showpacs,superscriptaddress,nofootinbib,floatfix,10pt]{revtex4-2}
\pdfoutput=1
\usepackage{textcomp}
\usepackage[english]{babel}
\usepackage{amsmath,amssymb,amsbsy,booktabs}
\usepackage{bm}
\usepackage{color}
\usepackage{array}
\usepackage{amstext}
\usepackage{graphicx}
\usepackage{amsfonts}
\usepackage{bm}
\usepackage{dcolumn}
\usepackage{rotating}
\usepackage{epstopdf}
\usepackage{esint}
\usepackage{url}
\usepackage[colorlinks=true,
linkcolor=blue,
filecolor=blue,
anchorcolor=blue,
urlcolor=blue,
citecolor=blue
]{hyperref}

\usepackage{array}
\usepackage{booktabs}
\usepackage{multirow}
\newcommand{\tabincell}[2]{\begin{tabular}{@{}#1@{}}#2\end{tabular}}

\usepackage[utf8]{inputenc}

\usepackage{xcolor,colortbl}

\usepackage[T1]{fontenc} 

\allowdisplaybreaks

\begin{document}

\title {\bf \boldmath Prospects for discovering new physics in charm sector through \\[0.1cm] low-energy scattering processes $e^-p\to e^- (\mu^-)\Lambda_c$}

\author{Li-Fen Lai}
\email{lailifen@mails.ccnu.edu.cn}
\affiliation{Institute of Particle Physics and Key Laboratory of Quark and Lepton Physics~(MOE),\\
	Central China Normal University, Wuhan, Hubei 430079, China}

\author{Xin-Qiang Li}
\email{xqli@mail.ccnu.edu.cn}
\affiliation{Institute of Particle Physics and Key Laboratory of Quark and Lepton Physics~(MOE),\\
	Central China Normal University, Wuhan, Hubei 430079, China}

\author{Xin-Shuai Yan}
\email{xinshuai@mail.ccnu.edu.cn}
\affiliation{Institute of Particle Physics and Key Laboratory of Quark and Lepton Physics~(MOE),\\
	Central China Normal University, Wuhan, Hubei 430079, China}

\author{Ya-Dong Yang}
\email{yangyd@mail.ccnu.edu.cn}
\affiliation{Institute of Particle Physics and Key Laboratory of Quark and Lepton Physics~(MOE),\\
	Central China Normal University, Wuhan, Hubei 430079, China}
\affiliation{School of Physics and Microelectronics, Zhengzhou University, Zhengzhou, Henan 450001, China}
	
\begin{abstract}
We investigate the potential for discovering new physics through the low-energy scattering processes $e^-p\to e^-(\mu^-)\Lambda_c$, which could be accessible in the forthcoming $ep$ scattering experiments once adapted to our proposed setups. In the framework of a general low-energy effective Lagrangian, we demonstrate that, compared with the conventional flavor-changing neutral-current weak decays of charmed hadrons and the dilepton productions at high-energy colliders, the low-energy scattering processes can provide more competitive potentials for hunting the underlying new physics. In some specific leptoquark models, we also show that promising event rates can be expected for both the scattering processes, and point out a possible way to distinguish the experimental signals due to the scalar leptoquarks from those of the vector ones.   
\end{abstract}


\maketitle

\section{Introduction} 

The absence of flavor-changing neutral current (FCNC) at tree level in the Standard Model (SM) makes these processes very sensitive to new physics (NP) beyond the SM. The FCNC processes in the beauty sector are known to be an ideal ground for NP searches due to their good exposure to the short-distance effects~\cite{Blake:2015tda,Isidori:2010kg}. However, similar processes in the charm sector are generally hindered by the pollution of long-distance hadronic contributions, which obscures the short-distance NP contributions. Therefore, to access the short-distance physics in the charm sector, \textit{e.g.}, in the semileptonic decays of charmed hadrons, one usually selects particular kinematic phase spaces, in which 
the NP-induced rates are much larger than those due to the hadronic resonance background~\cite{Artuso:2008vf,Burdman:2001tf,Gisbert:2020vjx}.

For a more thorough analysis, however, one may prefer to look at all the kinematically available regions in these rare decays, and thus a special treatment of the long-distance hadronic contributions becomes necessary. This can be achieved by using, \textit{e.g.}, a Breit-Wigner shape to model the resonance effect and then shifting it to the short-distance 
Wilson coefficients~\cite{Burdman:2001tf,Gisbert:2020vjx,deBoer:2015boa,deBoer:2015boa,Fajfer:2015mia,Fajfer:2005ke}.\footnote{For an alternative approach to this problem, see \textit{e.g.} Refs.~\cite{Feldmann:2017izn,Bharucha:2020eup}, where the hadronic resonance contributions are obtained through a dispersion relation, modeling the spectral functions as towers of Regge-like resonance in each channel and imposing the partonic behavior in the deep Euclidean.} Similar treatments of the hadronic resonance contributions in other 
charm decays can be found in Ref.~\cite{Burdman:2001tf}. Nevertheless, it is known that such treatments leave the nuisance parameters introduced to describe the resonance effects undetermined, 
due to the lack of experimental data~\cite{deBoer:2015boa, Gisbert:2020vjx, Fajfer:2005ke, Burdman:2001tf, deBoer:2015boa,Fajfer:2015mia}.

Crossing symmetry offers a way to avoid the long-distance pollution without invoking the kinematic cuts. To this end, instead of the conventional charmed-hadron weak decays mediated by the partonic-level $c\to u \ell^- \ell^+$ transitions, we consider the low-energy scattering process induced by the transition $\ell u \to \ell c$, whose amplitude is unambiguously connected to that of the former because of the crossing symmetry. Furthermore, due to its different kinematics relative to the rare FCNC decays, such a process is free from the long-distance pollution. To be more specific, let us consider the process of a low-energy electron scattered from a proton target and producing a $\Lambda_c$ baryon, \textit{i.e.}, $e^-p\to e^-\Lambda_c$, which can be considered as the FCNC production of the baryon $\Lambda_c$, analogous to, \textit{e.g.}, the electroproduction of the baryon $\Delta(1236)$ through $ep$ scattering~\cite{Cahn:1977uu}. It is also noted that the high-precision data of the Hadron-Electron Ring Accelerator has been recently used to constrain the NP contributions to the 
$eeqq$ contact interactions~\cite{ZEUS:2019cou}. The underlying processes involved are essentially the partonic-level $e q \to e q$ scatterings at high energy. However, only the interactions with the same quark flavors have been considered~\cite{ZEUS:2019cou}.   
 
Recently, thanks to the high luminosity accumulated at the Large Hadron Collider (LHC), competitive constraints on the short-distance NP effects in the charm sector can also be 
set by recasting resonant searches in the high-$p_T$ invariant 
mass tails of dilepton in the processes $pp\to \ell^-\ell^+$~\cite{Fuentes-Martin:2020lea}. Being essentially the $q\bar{q}\to \ell^-\ell^+$ scatterings realized in the high-energy regions, these processes are free from the long-distance pollution as well. As demonstrated in Ref.~\cite{Fuentes-Martin:2020lea}, there exhibits an interesting and pronounced
complementarity between the constraints from $pp\to \ell^-\ell^+$ high-$p_T$ invariant mass tails and those from the rare charm decays.

In this paper, we will show that, benefiting from the development of electron beams, 
particularly those designed for light dark-matter searches through electron fixed-target experiments~\cite{Abrahamyan:2011gv,Essig:2010xa,Essig:2013lka,Allison:2014tpu}, 
the FCNC scattering process $e^-p\to e^- \Lambda_c$, once measured with a proper experimental setup, could be competitive with the conventional charmed-hadron weak decays and the high-$p_T$ invariant mass tails, in constraining the short-distance NP effects in the charm sector.

Presumably, the simplest way to access the short-distance NP effects while 
without worrying about the long-distance pollution in charmed-hadron weak decays is to consider the null test observables, \textit{i.e.,} the observables that are very small in the SM due to approximate symmetries and/or parametric suppression~\cite{Fajfer:2015mia,DeBoer:2018pdx,Bause:2019vpr,Bause:2020xzj}. In this respect, the lepton flavor-violating (LFV) FCNC process is a sound candidate, since it contains practically no SM contribution due to the smallness of neutrino masses. This kind of process has been extensively studied in rare charmed-meson weak decays~\cite{deBoer:2015boa,DeBoer:2018pdx,Bause:2019vpr,Bause:2020xzj,Aaij:2015qmj,Lees:2011hb,LHCb:2020car,BaBar:2020faa}. Searches for the rare FCNC decays of charmed baryons, such as  $\Lambda_c$ and $\Sigma_c$, have also been conducted experimentally~\cite{Lees:2011hb, LHCb:2017yqf, E653:1995rpz}, while the corresponding detailed theoretical studies are still developing~\cite{Golz:2021imq}. The LFV FCNC processes have also been explored by analyzing the high-$p_T$ dilepton invariant mass tails in the processes $pp\to \ell \ell'$~\cite{Angelescu:2020uug}. Similar to the lepton flavor-conserving (LFC) case discussed above, the high-$p_T$ invariant mass tails are generally found to offer better insights into the short-distance NP effects. However, to have a deep understanding of the whole picture, 
they must also be complemented with the rare charm decays~\cite{Angelescu:2020uug}. 
  
In this paper, we will also consider the process of a low-energy electron scattered from a proton target and producing a $\Lambda_c$ baryon and a muon, \textit{i.e.}, $e^-p\to \mu^-\Lambda_c$. Interestingly enough, such a LFV scattering process, together with the LFC one discussed above, can be simultaneously accessible with one experimental setup. Based on the setup, we will show that our proposal of the low-energy LFV $ep$ scattering experiment can yield at least comparable constraints with respect to those obtained at the high-energy collider~\cite{Angelescu:2020uug}.

Among the possible new degrees of freedom that can mediate both the LFC and LFV scattering processes at tree level, we will focus on the leptoquark (LQ) mediators, which are expected to exist in several extensions of the SM, such as in the Pati-Salam model~\cite{Pati:1974yy} and in the grand unification theories~\cite{Georgi:1974sy,Fritzsch:1974nn}. These hypothetical particles can convert a quark into a lepton and vice versa and, due to such a distinctive character, have very rich phenomenology in precision experiments and at particle colliders~\cite{Dorsner:2016wpm,Buchmuller:1986zs}. Particularly, several anomalies observed recently in charged- and neutral-current $B$-meson weak decays have attracted extensive studies of the LQ interactions, due to their potentials for explaining the anomalies simultaneously (see, \textit{e.g.}, Refs.~\cite{Alonso:2015sja,Bauer:2015knc,Barbieri:2015yvd,Becirevic:2016yqi,Sahoo:2016pet,Crivellin:2017zlb,Alok:2017jaf,Buttazzo:2017ixm,Bordone:2018nbg,Becirevic:2018afm,Kumar:2018kmr,Angelescu:2018tyl,Cornella:2019hct,Crivellin:2019dwb,Babu:2020hun,Angelescu:2021lln,Cornella:2021sby}). While most of these analyses have been focusing on the LQ couplings to down-type quarks, we will concentrate on the LQ interactions involving the light leptons and up-type quarks. For completeness, we consider both scalar and vector LQs, even though the vector ones are more sensitive to the ultraviolet complete models, which may in turn render the obtained limits on vector LQ couplings less robust~\cite{Mandal:2019gff}. With the selected experimental setup, together with the constraints from high-$p_T$ invariant mass tails and the rare charm decays, we will show that the observation of both the LFC and LFV scattering processes mediated by the LQs can be expected.  

The paper is organized as follows. In Sec.~\ref{sec:models}, we start with a recapitulation of the generic LQ-fermion interactions that respect the SM gauge symmetry. To avoid the stringent constraints set by the null results of experimental searches for proton decays, we exclude the LQs that can induce tree-level proton decays. We then present the general effective Lagrangian that can mediate the two scattering processes. In the established theoretical framework, we firstly consider in Sec.~\ref{sec:LFC_FCNC} the LFC $e^-p\to e^- \Lambda_c$ and then in Sec.~\ref{sec:LFV_FCNC} the LFV $e^-p\to \mu^- \Lambda_c$ scattering process, including both the kinematics involved and the proper experimental setups. After revisiting the currently existing experimental constraints, we evaluate the prospect for discovering the potential LQ effects through the low-energy scattering processes in various aspects. Our conclusions are finally made in Sec.~\ref{sec:con}. For convenience, the helicity-based definitions for $\Lambda_c\to p$, the cross section and kinematics, as well as the amplitude squared of the LFC (LFV) scattering process are given in Appendices \ref{appendix:form factor}, \ref{appendix:cross section}, and \ref{appendix:amplitude}, respectively.

\section{Theoretical framework}
\label{sec:models}

\subsection{Leptoquarks}

We start with a brief summary of the LQ interactions with the SM fermions. Based on their different representations under the SM gauge group $\text{SU(3)}_C\times \text{SU}(2)_{L}\times \text{U}(1)_{Y}$, both scalar and vector LQs can be classified into five different categories. Following the notation used commonly for the LQs in the literature~\cite{Buchmuller:1986zs,Dorsner:2016wpm}, we present schematically 
in Table~\ref{tab:LQ} their interactions with the SM fermions, where the left-handed lepton (quark) doublets are denoted as $L_L^i=(\nu_L^i,\ell_L^i)^T$ ($Q_L^i=(u_L^i,d_L^i)^T$), while the right-handed up-(down-)type quark and lepton singlets as $u_R^i$ ($d_R^i$) and $e_R^i$, respectively. Note that, for simplicity, neither the coupling constants nor the Hermitian conjugation are explicitly shown. 

\begin{table}[ht]
		\renewcommand*{\arraystretch}{1.3}
		\tabcolsep=0.13cm
		\centering
		\begin{tabular}{|cc|cc|}
			\hline \hline
			Scalar LQ & Representation & Vector LQ & Representation
			\\ \hline \rowcolor{lightgray}
			\tabincell{c}{$S_1 Q_L L_L$ \\ $S_1u_Re_R$} & $(\bar{3}, 1, 1/3)$ & 
			\tabincell{c}{$U_{1\mu}\bar{Q}_L\gamma^{\mu}L_L$ \\ $U_{1\mu}\bar{d}_R\gamma^{\mu}e_R$} & $(3, 1, 2/3)$\\ 
			\tabincell{c}{$R_2\bar{u}_R L_L$ \\ $R_2\bar{Q}_L e_R $} & $(3, 2, 7/6)$ & \tabincell{c}{$V_{2\mu}d_R\gamma^{\mu} L_L$ \\   $V_{2\mu}Q_L\gamma^{\mu}e_R$} & $(\bar{3}, 2, 5/6)$\\ \rowcolor{lightgray}
			$S_3 Q_L L_L$ & $(\bar{3}, 3, 1/3)$ & $U_{3\mu} \bar{Q}_L\gamma^{\mu}L_L$ & $(3, 3, 2/3)$ \\ 
			$\tilde{S}_1 d_R e_R$ & $(\bar{3}, 1, 4/3)$ & $\tilde{U}_{1\mu} \bar{u}_R\gamma^{\mu}e_R$ & $(3, 1, 5/3)$ \\ \rowcolor{lightgray}
			$\tilde{R}_2 \bar{d}_R L_L$ & $(3, 2, 1/6)$ & $\tilde{V}_{2\mu}u_R\gamma^{\mu}L_L$ & $(\bar{3}, 2, -1/6)$
			\\
			\hline\hline
		\end{tabular}
		\caption{Scalar and vector LQ interactions with the SM fermions, as well as their representations under the SM gauge group $\text{SU(3)}_C\times \text{SU}(2)_{L}\times \text{U}(1)_{Y}$. Our convention for the hypercharge $Y$ is given by $Q_{\text{em}}=T_3+Y$.} 
		\label{tab:LQ} 
\end{table} 

It is known that not all the LQs can be assigned definite baryon ($B$) and lepton ($L$) numbers, because some of them can also couple to two quarks, which would then mediate proton decays at tree level~\cite{Arnold:2013cva,Gardner:2018azu,Assad:2017iib}. Here, given the strong constraints on proton stability (or generally the $|\Delta B|=1$ processes)~\cite{Zyla:2020zbs}, we will focus only on the LQs that cannot mediate these processes at tree level. This reduces the number of LQs significantly, and leaves us with only one scalar $R_2$ and three vector $(U_{1},U_3,\tilde{U}_1)$ LQs. One can further reduce the number to one by forbidding dimension-five proton decays as well --- only $\tilde{U}_1$ would survive in this case~\cite{Assad:2017iib}. However, since the dimension-five proton decay operators can be conveniently eliminated by embedding the vector LQs into UV complete models or by imposing some additional symmetry (such as a gauged $\text{U(1)}_{B-L}$) for the scalar case~\cite{Arnold:2013cva,Assad:2017iib}, we will keep the LQs that can induce dimension-five proton decays.

Since the vector LQ $U_1$ cannot mediate the scattering processes $e^-p\to e^- (\mu^-)\Lambda_c$ at tree level, we will concentrate only on the remaining three LQs $(R_2,U_3,\tilde{U}_1)$, with their relevant interactions fleshed out, respectively, as
\begin{align}
\mathcal{L}_{R_2}&\supset R^{\frac{5}{3}}_2\left[(\lambda^S_2)_{ij}\bar{u}^i_{R}e^j_L+(\lambda^{\prime S}_2)_{ij}\bar{u}^i_{L}e^j_R\right]+\text{H.c.}, \label{eq:R2}\\[0.2cm]
\mathcal{L}_{U_3}&\supset U^{\frac{5}{3}}_{3\mu}(\lambda^V_3)_{ij} \bar{u}^i_{L}\gamma^{\mu} e^j_L+\text{H.c.}, \label{eq:U3}\\[0.2cm]
\mathcal{L}_{\tilde{U}_1}&\supset \tilde{U}^{\frac{5}{3}}_{1\mu}(\lambda^V_{\tilde{1}})_{ij} \bar{u}^i_{R}\gamma^{\mu} e^j_R+\text{H.c.}, \label{eq:tildeU1} 
\end{align}
where we have adopted the four-component spinors for all the fermions such that, \textit{e.g.}, $e_{R,L}=P_{R,L}e$, with $P_{R,L}=(1\pm\gamma^5)/2$. Several necessary explanations are in order. First, we have assigned in Eq.~\eqref{eq:R2} two different coupling constants $(\lambda^S_2)_{ij}$ and $(\lambda^{\prime S}_2)_{ij}$, with $i,j=1,2,3$ denoting the generation indices, to characterize the two different interactions of $R_2$ with the SM fermions. Note that there exists a minus sign difference between the couplings $\lambda^S_2$ in Eq.~\eqref{eq:R2} and $y^{RL}_2$ defined in Ref.~\cite{Dorsner:2016wpm}. In addition, for the LQ $U_3$, our definition of the coupling $\lambda^V_3$ in Eq.~\eqref{eq:U3} differs by a factor $\sqrt{2}$ from $x^{LL}_3$ introduced in Ref.~\cite{Dorsner:2016wpm}. Second, the electric charges of the LQ components are characterized by the rational superscript in each of the LQs, with the convention $Q_{\text{em}}=T_3+Y$. Finally, all the fermion fields in Eqs.~\eqref{eq:R2}--\eqref{eq:tildeU1} have been given in their mass-eigenstate basis, which, in our convention, coincides with the flavor basis of the left-handed up-type quarks and charged leptons. With such a convention, the left-handed down-type quark and neutrino flavor states can be transformed into their respective mass eigenstates through $d_L\to Vd_L$ and $\nu_L\to U\nu_L$, where $V$ and $U$ denote the Cabibbo-Kobayashi-Maskawa and the Pontecorvo-Maki-Nakagawa-Sakata matrix, respectively.    

\subsection{Effective Lagrangian}

The general effective Lagrangian responsible for the process $\ell u \to \ell (\ell') c$ can be written as
\begin{align}
\mathcal{L}_{\text{eff}}=\mathcal{L}^{\text{SM}}_{\text{eff}}+\mathcal{L}^{\text{LQ}}_{\text{eff}},
\end{align}
where $\mathcal{L}^{\text{SM}}_{\text{eff}}$ and $\mathcal{L}^{\text{LQ}}_{\text{eff}}$ represent the SM and the LQ contribution, respectively. In contrast to the charmed-hadron weak decays, to which both the short- and long-distance effects from the SM can contribute, only the short-distance ones play a part in the scattering processes. However, due to the Glashow-Iliopoulos-Maiani mechanism, the SM short-distance effects are strongly suppressed~\cite{Burdman:2001tf,Gisbert:2020vjx,Paul:2011ar,Cappiello:2012vg,deBoer:2015boa,Fajfer:2015mia,
Bause:2019vpr,Fajfer:2015zea,Meinel:2017ggx,Azizi:2010zzb,Sirvanli:2016wnr}, and thus we can safely neglect the contribution from $\mathcal{L}^{\text{SM}}_{\text{eff}}$, when discussing the scattering processes.

The general effective Lagrangian $\mathcal{L}^{\text{LQ}}_{\text{eff}}$ induced by tree-level exchanges of LQs is given by~\cite{Mandal:2019gff} 
\begin{align}
\mathcal{L}^{\text{LQ}}_{\text{eff}}
=&\sum_{i,j,m,n}\Big\{ [g^{LL}_{V}]^{ij,mn}(\bar{\ell}^i_L\gamma_{\mu}\ell^j_{L}) (\bar{q}^m_L\gamma^{\mu}q^n_{L})\nonumber \\[0.15cm]
&+ [g^{LR}_{V}]^{ij,mn}(\bar{\ell}^i_L\gamma_{\mu}\ell^j_{L}) (\bar{q}^m_R\gamma^{\mu}q^n_{R}) \nonumber \\[0.15cm]
&+ [g^{RL}_{V}]^{ij,mn}(\bar{\ell}^i_R\gamma_{\mu}\ell^j_{R}) (\bar{q}^m_L\gamma^{\mu}q^n_{L})\nonumber \\[0.15cm] 
&+ [g^{RR}_{V}]^{ij,mn}(\bar{\ell}^i_R\gamma_{\mu}\ell^j_{R}) (\bar{q}^m_R\gamma^{\mu}q^n_{R}) \nonumber \\[0.15cm]
&+ [g^{L}_{T}]^{ij,mn}(\bar{\ell}^i_R\sigma^{\mu\nu}\ell^j_{L}) (\bar{q}_R^m\sigma_{\mu\nu} q_L^n)\nonumber \\[0.15cm] 
&+ [g^{R}_{T}]^{ij,mn}(\bar{\ell}^i_L\sigma^{\mu\nu}\ell^j_{R}) (\bar{q}_L^m\sigma_{\mu\nu}q_R^n) \nonumber \\[0.15cm]
&+ [g^{L}_{S}]^{ij,mn}(\bar{\ell}^i_R\ell^j_{L}) (\bar{q}^m_Rq^n_{L})\nonumber \\[0.15cm]
&+ [g^{R}_{S}]^{ij,mn}(\bar{\ell}^i_L\ell^j_{R}) (\bar{q}^m_Lq^n_{R})\Big\}, \label{eq:Lag_LQ}
\end{align}
where $i, j$ and $m, n$ represent the flavor indices of leptons and quarks, respectively. The effective Wilson coefficients (WCs) $g$ resulting from different LQs will be explicitly determined case by case, and only the non-vanishing ones will be shown hereafter. 

After integrating out the scalar LQ $R_2$, together with proper chiral Fierz transformations (see, \textit{e.g.}, Ref.~\cite{Nishi:2004st}) of the resulting four-fermion operators to the ones given by Eq.~\eqref{eq:Lag_LQ}, we get the following non-vanishing WCs:
\begin{align}
[\hat{g}^{L}_{S}]^{ij,mn}&=4 [\hat{g}^{L}_{T}]^{ij,mn}
=-(\lambda'^S_2)^*_{in}(\lambda^S_2)_{jm}, \nonumber \\[0.2cm] 
[\hat{g}^{R}_{S}]^{ij,mn}&=4 [\hat{g}^{R}_{T}]^{ij,mn}
=-(\lambda^S_2)^*_{in}(\lambda'^S_2)_{jm}, \nonumber \\[0.2cm] 
[\hat{g}^{LR}_{V}]^{ij,mn}&=-(\lambda^S_2)^*_{in}(\lambda^S_2)_{jm}, \nonumber \\[0.2cm] 
[\hat{g}^{RL}_{V}]^{ij,mn}&=-(\lambda'^S_2)^*_{in}(\lambda'^S_2)_{jm},\label{eq:g_denfi_1}
\end{align}
where we have defined $g\equiv \omega \hat{g}$ with the common factor $\omega\equiv 1/(2M^2)$. For simplicity, we will assume that all the three LQs share the same mass $M$, which is of course not necessary the case in nature. For tree-level exchanges of the vector LQs $U_3$ and $\tilde{U}_1$, on the other hand, only one non-vanishing WC survives in each case. Following the same procedure, we obtain  
\begin{align}
 [\hat{g}^{LL}_{V}]^{ij,mn}&=-2(\lambda^V_3)^*_{in}(\lambda^V_3)_{jm}, \label{eq:g_denfi_2} \\[0.2cm]
 [\hat{g}^{RR}_{V}]^{ij,mn}&=-2(\lambda^V_{\tilde{1}})^*_{in}(\lambda^V_{\tilde{1}})_{jm},\label{eq:g_denfi_3}
\end{align}
for $U_3$ and $\tilde{U}_1$, respectively. 

The WCs and their relations given by Eqs.~\eqref{eq:g_denfi_1}--\eqref{eq:g_denfi_3} hold at the matching scale $\mu=M$. To connect the LQ coupling constants $\lambda$~($\lambda'$) to the low-energy scattering processes, we must use the renormalization group (RG) equation to evolve them to the corresponding low-energy scale. Since large mixings of the tensor operators into the scalar ones can arise due to QED and electroweak (EW) one-loop effects~\cite{Gonzalez-Alonso:2017iyc,Aebischer:2017gaw}, we take account of both QCD and EW/QED effects. To be specific, we firstly match at the NP scale $\mu=M$ the general effective Lagrangian given by Eq.~\eqref{eq:Lag_LQ} to that of the Standard Model Effective Field Theory (SMEFT), in which the four-fermion operators are defined in the Warsaw basis (see Table~3 in Ref.~\cite{Grzadkowski:2010es}), and then perform the RG running from $\mu=M$ down to the EW scale ($\mu=m_Z$). Here we take $M=1$~TeV as the benchmark for the LQ mass, since direct searches for the LQs at LHC have already pushed the lower bounds to such an energy scale~\cite{CMS:2012lqv,Aad:2015caa,Sirunyan:2018ruf,CMS:2020wzx}.\footnote{Note that the lower mass bounds for vector LQs have been pushed roughly up to $1.8$~TeV~\cite{CMS:2020wzx}. Nonetheless, we choose here $1$~TeV for both scalar and vector LQs as a simple demonstration.} Finally, we match the four-fermion SMEFT operators to the low-energy ones given at the EW scale, and continue the RG running down to the characteristic scale $\mu=2$~GeV. Taking into account both the one-loop QCD and EW/QED effects~\cite{Jenkins:2017dyc,Gonzalez-Alonso:2017iyc}, we obtain numerically
\begin{align}
g^{\chi}_{S}(2\,\text{GeV})&\approx 2.0\, g^{\chi}_S(1\,\text{TeV})-0.5\, g^{\chi}_T(1\,\text{TeV}), \nonumber \\[0.2cm]
g^{\chi}_T(2\,\text{GeV})&\approx 0.8\, g^{\chi}_T(1\,\text{TeV}),\label{eq:RG_R2_high}
\end{align} 
where $\chi=L, R$. We neglect the RG running effects of the vector operators, since these operators do not get renormalized under QCD while their RG running effects under EW/QED are only at the percent level. Note that our result in Eq.~\eqref{eq:RG_R2_high} is slightly different from that in Eq.~(6.4) of Ref.~\cite{Fuentes-Martin:2020lea}, which results from the two- and three-loop QCD effects~\cite{Gonzalez-Alonso:2017iyc} that have been neglected here.

Together with the result in Eq.~\eqref{eq:RG_R2_high}, the scalar-tensor WC relations, $\hat{g}^{\chi}_{S}(1\,\text{TeV})=4 \hat{g}^{\chi}_T(1\,\text{TeV})$, in Eq.~\eqref{eq:g_denfi_1} would be modified as 
\begin{align}
g^{\chi}_{S}(2\,\text{GeV})\approx 9.4\, g^{\chi}_T(2\,\text{GeV}), \label{eq:RG_R2}
\end{align}
at the scale $\mu=2$~GeV for $R_2$, the only scalar LQ that can generate non-vanishing tensor and scalar effective operators considered in this work.  

The general effective Lagrangian introduced in Eq.~\eqref{eq:Lag_LQ} is often presented in another operator basis (see, \textit{e.g.}, Refs.~\cite{deBoer:2015boa,Bause:2019vpr}). For the LFC case, the conversion relations between the WCs defined in these two bases are given by 
\begin{align}
C_{9,10}&=\frac{\sqrt{2}\pi}{2G_F\alpha_e}\left(g^{RL}_V\pm g^{LL}_V\right),\quad C_{S,P}=\frac{\sqrt{2}\pi}{2G_F\alpha_e}g^{R}_S,\nonumber \\
C'_{9,10}&=\frac{\sqrt{2}\pi}{2G_F\alpha_e}\left(g^{RR}_V\pm g^{LR}_V\right),\quad  C'_{S,P}=\pm\frac{\sqrt{2}\pi}{2G_F\alpha_e}g^{L}_S, \nonumber \\
C_{T,T_5}&=\frac{\sqrt{2}\pi}{G_F\alpha_e}\left(g^{R}_T\pm g^{L}_T\right), \label{eq:relation}
\end{align}
where $G_F$ is the Fermi constant and the fine-structure constant $\alpha_e\equiv e^2/4\pi$. For the LFV case, the conversion relations in Eq.~\eqref{eq:relation} still hold and, to be distinguished from the LFC case, all the $C_i$~($C'_i$) will be replaced by $K_i$~($K'_i$). Note that, for convenience, we will denote the $g(2\,\text{GeV})$ simply by $g$ hereafter.
 
\subsection{Initial- and final-state hadronic effects}

Before diving into explicit analyses of the two scattering processes $e^-p\to e^-\Lambda_c$ and $e^-p\to \mu^-\Lambda_c$, we now discuss how to characterize the initial- and final-state hadronic effects. Since the energy scale of the two processes is in a few GeV region, it is unnecessary to invoke the parton distribution functions, while the form-factor description is sufficient. In fact, similar approaches have been adopted in the asymmetrical electroproduction of the baryon $\Delta(1236)$ by polarized $ep$ scattering~\cite{Cahn:1977uu,Nath:1979qe}, in parametrizing the parity violation 
in electron-nucleon scattering (see, \textit{e.g.}, Ref.~\cite{Souder:2015mlu}), and recently in evaluating the cross section of inverse beta decay~\cite{Ankowski:2016oyj,Ankowski:2019tbc}.

We will adopt the helicity-based definition of the $\Lambda_c\to p$ form factors~\cite{Feldmann:2011xf,Meinel:2017ggx}. For details, we refer the readers to Appendix~\ref{appendix:form factor}. The form factors can be calculated in quark models~\cite{Ivanov:1996fj,Pervin:2005ve,Gutsche:2014zna,Faustov:2016yza}, by QCD light-cone sum rules~\cite{Li:2016qai}, or through Lattice QCD (LQCD)~\cite{Meinel:2017ggx}. We will follow the following two criteria to select the proper methods: (i) the form-factor parametrization must be analytic in the complex $q^2$ plane, with cut along the real axis for $q^2\geq t_+$ with $t_+=(m_{\Lambda_c}+m_p)^2$; (ii) an error estimation for the form factors can be provided. The first criterion is crucial, because the kinematic range of a decay process is different from that of a scattering process. For example, the kinematic range of the semileptonic decay $\Lambda_c\to p \ell^- \ell^+$ is given by $0\leq q^2\leq t_-$ 
with $t_-=(m_{\Lambda_c}-m_p)^2$, whereas that of the low-energy scattering process $e^-p\to e^-\Lambda_c$ is restricted to $q^2<0$. Consequently, to warrant the application of form-factor descriptions that are conventional for the charmed-hadron weak decays to the low-energy scattering processes, the form-factor parametrization must possess analyticity in the proper $q^2$ range. Remarkably, such a parametrization scheme already exists. It was initially proposed to parametrize the $B\to \pi$ vector form factor in Ref.~\cite{Bourrely:2008za}, and has been recently utilized in the LQCD calculation of the $\Lambda_c\to N$ (nucleon) form factors in Ref.~\cite{Meinel:2017ggx}.  Given the LQCD calculation can also provide an error estimation, we will thus adopt their latest results~\cite{Meinel:2017ggx} in this work.  

\section{\boldmath LFC FCNC process $e^-p\to e^-\Lambda_c$}
\label{sec:LFC_FCNC}

In this section, we evaluate the prospect for discovering the LFC process $e^-p\to e^-\Lambda_c$ mediated by the survived LQs through low-energy scattering experiments. In this context, based on the general effective Lagrangian given by Eq.~\eqref{eq:Lag_LQ} at $\mu=2$~GeV, we briefly discuss its kinematics and cross section. The benefit of working in the framework of low-energy $\mathcal{L}_{\text{eff}}$ here is twofold. First, the connection of the low-energy scattering process to the high-energy LQ models can always be established through the relations in Eqs.~\eqref{eq:g_denfi_1}--\eqref{eq:g_denfi_3}. Second, the low-energy result can be easily generalized to other ultraviolet models. After building a suitable experiment setup, we will show that, compared with the charmed-hadron weak decays and the high-$p_T$ dilepton invariant mass tails, the low-energy scattering process can 
provide a competitive insight into the NP effects in charm sector. Finally, taking into account the constraints from charmed-hadron weak decays and high-$p_T$ dilepton invariant mass tails, we provide an event-rate estimation for the LFC scattering experiment in the survived LQ models.  

\subsection{Cross section and kinematics}

The differential cross section of the scattering process $e^-(k)+p(P)\to e^-(k')+\Lambda_c(P')$, with $P=(m_p, 0)$, $P'=(E_{\Lambda_c}, \pmb{p}')$, $k=(E, \pmb{k})$, and $k'=(E', \pmb{k}')$, is given by 
\begin{align}
d\sigma=&\frac{1}{4[(P\cdot k)^2-m^2_em^2_p]^{1/2}}\frac{d^3\pmb{k}'}{(2\pi)^3}\frac{1}{2E'}\frac{d^3\pmb{p}'}{(2\pi)^3}\frac{1}{2E_{\Lambda_c}}\nonumber \\[0.15cm]
&\times \overline{|\mathcal{M}|}^2(2\pi)^4\delta^4(P+k-P'-k'),\label{eq:LFC_diff_cross}
\end{align}
where the amplitude squared $\overline{|\mathcal{M}|}^2$ is obtained by averaging over the initial- and summing up the final-state spins, with more details elaborated in Appendix~\ref{appendix:cross section}. Contrary to the semileptonic charmed-hadron decays, the kinematics of the scattering process is now bounded by 
\begin{align}
	\frac{2E(m_{\Lambda_c}^2-m_{p}^2-2m_{p}E)}{m_{p}+2E}\leq q^2 \leq 0, \label{eq:LFC_Q2_range}
\end{align}
which indicates that the electron beam energy $E$ must satisfy condition 
\begin{align}
E\geq (m_{\Lambda_c}^2-m_{p}^2)/(2m_{p})\label{eq:LFC_E_cond}.
\end{align}

\subsection{Relevant experimental constraints}
\label{subsection:LFC_other_constraints}

We here revisit briefly the most relevant and stringent constraints on the effective coefficients $[g]^{ee,cu}$ from the charmed-hadron weak decays and the high-$p_T$ dilepton invariant mass tails.

\begin{itemize}
	\item \textit{Constraint from $D^0\to e^+e^-$}
\end{itemize}

The expression of the branching ratio of $D^0\to \ell^+\ell^-$ decay can be found, \textit{e.g.}, in Refs.~\cite{deBoer:2015boa,Bause:2019vpr}, where it is usually formulated in terms of $C$ and $C'$. Using the experimental limit on the branching fraction of $D^0\to e^+e^-$, $\mathcal{B}(D^0\to e^+e^-)<7.9\times 10^{-8}$ at $90\%$ confidence level (C.L.)~\cite{Petric:2010yt}, and neglecting the SM contributions as in  Refs.~\cite{deBoer:2015boa,Bause:2019vpr}, we obtain 
\begin{align}
	|C_S-C'_S|^2+|C_P-C'_P|^2 \lesssim 0.062,
\end{align}
where terms proportional to $C_{10}$ and $C'_{10}$ have been neglected due to the tiny mass ratio $m_em_c/m_D^2$. Using the relations in Eq.~\eqref{eq:relation}, we can rewrite the constraint as 
\begin{align}
	\frac{\pi^2}{G^2_F\alpha^2_e}\left(|g^R_S|^2+|g^L_S|^2\right)\lesssim 0.062.\label{eq:Dee}
\end{align}

\begin{itemize}
	\item \textit{Constraints from $D^+\to \pi^+ e^+e^-$ and $\Lambda_c\to p e^+e^-$}
\end{itemize}

The differential decay distribution of $D\to P \ell^+\ell^-$ with $P$ standing for a pseudoscalar meson has been discussed, \textit{e.g.}, in Refs.~\cite{deBoer:2015boa,Bause:2019vpr}.\footnote{Note that the tensor matrix element in Ref.~\cite{Bause:2019vpr} is normalized by $m_D+m_P$, rather than by $m_D$ as in Ref.~\cite{deBoer:2015boa}. Here, we adopt the convention used in Ref.~\cite{Bause:2019vpr}.} Using the latest lattice result of the $D\to \pi$ form factors~\cite{Lubicz:2017syv,Lubicz:2018rfs}, we update the constraint on the WCs $C_i$ and $C'_i$ set by the measurement of $D^+\to \pi^+ e^+e^-$ decay. Currently, the best limit on the branching fraction of $D^+\to \pi^+ e^+e^-$ is $\mathcal{B}(D^+\to \pi^+ e^+e^-)<1.1\times 10^{-6}$ in the full $q^2$ regions, \textit{i.e.}, $\sqrt{q^2}\in [0.2,0.95]$~GeV and 
$\sqrt{q^2}>1.05$~GeV, at $90\%$ C.L.~\cite{Lees:2011hb}, which yields the constraint
\begin{align}
	& 0.07\,|C_9+C'_9|^2+0.07\,|C_{10}+C'_{10}|^2+0.16\,|C_S+C'_S|^2\nonumber \\[0.15cm] 
	& +0.16\,|C_P+C'_P|^2+0.02\,|C_T|^2+0.02\,|C_{T5}|^2\lesssim 1.
\end{align}
This translates into the following constraint on $g$: 
\begin{align}
	\frac{\pi^2}{G^2_F\alpha^2_e}&\Big[0.16\,(|g^L_S|^2+|g^R_S|^2)+0.08\,(|g^L_T|^2+|g^R_T|^2)\nonumber \\[0.15cm] 
	&+0.07\,(|g^{RL}_V|^2+|g^{LR}_V|^2+|g^{RR}_V|^2+|g^{LL}_V|^2\nonumber \\[0.15cm]
	&+2.0\,\text{Re}[g^{RL}_Vg^{RR*}_V+g^{LL}_Vg^{LR*}_V])\Big]\lesssim 1.\label{eq:DPee}
\end{align}

In comparison with the $D$-meson weak decays, studies on the rare FCNC decays of $\Lambda_c$ baryon are much fewer. Currently, the only existing experimental constraint on  $\mathcal{B}(\Lambda_c\to p e^+e^-)$ is set by the BaBar Collaboration~\cite{Lees:2011hb}.
Theoretical studies of $\Lambda_c\to p e^+e^-$ exist as well~\cite{Azizi:2010zzb,Sirvanli:2016wnr,Golz:2021imq}. However, no direct constraints on the NP WCs have been set---the same observation also holds for the LFV charmed-hadron decays. Therefore, we will concentrate on the constraint from $D^+\to \pi^+ e^+e^-$. 

\begin{itemize}
	\item \textit{Constraint from high-$p_T$ dilepton mass tails}
\end{itemize}

Due to the high luminosity accumulated at the LHC, constraining NP through the analysis of the invariant mass tails of the dilepton in $pp(q\bar{q})\to e^{+}e^{-}$ processes at high $p_T$ becomes feasible. Recently, the analysis from the CMS Collaboration with 140~$\text{fb}^{-1}$ of $13$-TeV data~\cite{CMS:2019tbu} has been recast into the following constraints on $g$ at $\mu=2$~GeV and at $90\%$ C.L.~\cite{Fuentes-Martin:2020lea}: 
\begin{align}
	|g^i_V|\!\lesssim \! 1.9\frac{G_F\alpha_e}{\pi}, \; |g^{L,R}_S|\!\lesssim \! 4.7\frac{G_F\alpha_e}{\pi}, \; |g^{L,R}_T|\!\lesssim \! 0.76\frac{G_F\alpha_e}{\pi}, \label{eq:qqee}
\end{align} 
where $i=LL,RR,LR,RL$. Note that, due to the negligible fermion masses involved in these processes, contributions from the interference terms, such as $g^{RL}_Vg^{RR*}_V$ and $g^{L}_Sg^{R*}_T$, have been neglected~\cite{Fuentes-Martin:2020lea,Angelescu:2020uug}. 

\begin{table}[ht]
	\renewcommand\arraystretch{1.3} 
	\tabcolsep=0.085cm
	\centering
	\begin{tabular}[t]{|c|c|c|c|c|}
		\hline \hline
		Processes &$\big|g^{LL,RR}_V\big|^2$ &$\big|g^{LR,RL}_V\big|^2$ & $\big|g^{L,R}_S\big|^2$& $\big|g^{L,R}_T\big|^2$  \\ \hline
		$D^0\to e^+e^-$~\cite{Petric:2010yt} &$\backslash$  &$\backslash$  & 0.062& $\backslash$ \\
		$D^+\to \pi^+e^+e^-$~\cite{Lees:2011hb} & 14& 14& 6.3 & 13 \\
		$pp(q\bar{q})\to e^+e^-$~\cite{CMS:2019tbu} & 3.6& 3.6 & 22&0.57 \\
		$e^-p\to e^-\Lambda_c$ & 0.035& 0.083 & 0.17&0.0056 \\
		\hline \hline
	\end{tabular}
	\caption{Constraints on the WCs $g$ at $90\%$ C.L. from the LFC (semi)leptonic $D$-meson decays, the high-$p_T$ dilepton invariant mass tails, and the $e^-p\to e^-\Lambda_c$ scattering process in the framework of a general low-energy effective Lagrangian specified by Eq.~\eqref{eq:Lag_LQ}. Note that we have factored out the common factor $G^2_F\alpha^2_e/\pi^2$. The entries with ``$\backslash$'' mean that the processes in the first column put no constraints on the corresponding WCs. }
	\label{table:constraints_LFC} 
\end{table} 

We summarize in Table~\ref{table:constraints_LFC} the aforementioned constraints. It can be seen that the most stringent constraint on $g_S$ results from the measurement of $D^0\to e^+e^-$, which unfortunately sheds no light on $g_V$ and $g_T$. Such a deficiency can be remedied by the measurement of $D^+\to \pi^+ e^+e^-$ decay and the analysis of high-$p_T$ dilepton invariant mass tails in $pp(q\bar{q})\to e^+e^-$. Clearly, the latter can provide better constraints than do the semileptonic $D$-meson decay. Nonetheless, a complementarity between the $D$-meson decays and the high-$p_T$ dilepton invariant mass tails can indeed be established~\cite{Fuentes-Martin:2020lea}. 

The model-independent constraints obtained thus far can be translated into the LQ cases. To this end, we must take into account the RG running effects (see Eq.~\eqref{eq:RG_R2_high}). For the scalar LQ $R_2$, we obtain
\begin{align}
	&|g^{RL,LR}_V|^2 \lesssim  3.6 \left(\frac{G^2_F\alpha_e^2}{\pi^2}\right),\nonumber \\
	&|g^{L,R}_S|^2\simeq 88\,|g^{L,R}_T|^2 \lesssim 0.062 \left(\frac{G^2_F\alpha_e^2}{\pi^2}\right),\label{eq:const_x4}
\end{align}
where the first one comes from $pp(q\bar{q})\to e^+e^-$ and the other one from $D^0\to e^+e^-$. It can be seen that both the scalar and tensor operators are severely constrained. For the vector LQ $U_{3}$ ($\tilde{U}_1$), on the other hand, we get
\begin{align}
	|g^{LL(RR)}_V|^2\lesssim 3.6 \left(\frac{G^2_F\alpha_e^2}{\pi^2}\right). \label{eq:const_v23}
\end{align} 

\subsection{Experimental setup}

We consider the fixed-target scattering experiments, whose event rate $dN/dt$ is defined by 
\begin{align}
	\frac{dN}{dt}=\mathcal{L}\sigma=\phi \rho_T L \sigma,
\end{align}
where the luminosity $\mathcal{L}$ is given in units of $\text{cm}^{-2}\text{s}^{-1}$, and the incoming beam is characterized by the flux $\phi$ (\textit{i.e.} the number of electrons per second), which is connected to the intensity $I$ through the relation $\phi=I/|e|$, with the electron charge $e=-|e|$. The target number density and length are denoted by $\rho_T$ and $L$, respectively.\footnote{Note that $\rho_T=\rho/m_p$, where $\rho$ is the density of the proton target.}   

\begin{figure}[t]
	\centering
	\includegraphics[width=0.48\textwidth]{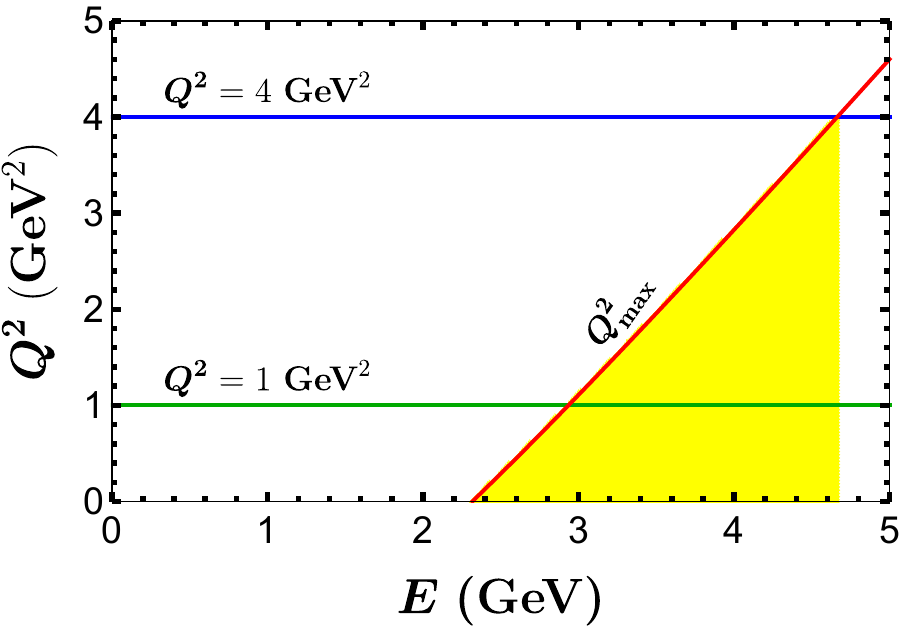}
	\caption{Criteria for selecting the electron beam energy $E$, where the red line denotes the $E-Q^2_{\max}$ relation given by Eq.~\eqref{eq:LFC_Q2_range}, the blue line represents the condition $Q^2\leq 4\,\text{GeV}^2$ required by our theoretical framework, while the green line corresponds to our benchmark scenario with $Q^2=1\,\text{GeV}^2$. The yellow region indicates the eligible $E$ with its corresponding $[Q^2_{\min},Q^2_{\max}]$.} 
	\label{fig:Eselection} 
\end{figure}

As demonstrated in Eq.~\eqref{eq:LFC_Q2_range}, the electron beam energy $E$ determines the maximal $Q^2$ ($Q^2=-q^2$) of the scattering process, which, in turn, implies that constraints on $Q^2_{\max}$ restrict the $E$ selection. As an explicit example, we have used the condition $Q^2_{\max}=Q^2_{\min}=0$ to obtain the minimal requirement for $E$ of the scattering process (see Eq.~\eqref{eq:LFC_E_cond}). This condition is also visualized in Fig.~\ref{fig:Eselection} by noting the intersection point of the $E$ axis and the red line that represents the $E$-$Q^2_{\max}$ relation. Besides the kinematic constraint on $Q^2_{\max}$, a limit from the theoretical point of view must be taken into account as well. As mentioned before, our analysis is carried out in the framework of $\mathcal{L}_{\text{eff}}$ given by Eq.~\eqref{eq:Lag_LQ} at the scale $\mu=2\,\text{GeV}$; to ensure the validity of our results presented below, we must require $Q^2_{\max}$ not to exceed $Q^2=4\,\text{GeV}^2$. Such a requirement, depicted by the blue line in Fig.~\ref{fig:Eselection}, indicates an upper bound $E\lesssim 4.65\,\text{GeV}$, provided that the observables one is interested in, such as the total cross section, involve $Q^2_{\max}$. Otherwise, $E$ is not bounded as above---one can always focus on the lower $Q^2$ region, even though a high $Q_{\max}^2$ is available due to a high $E$. In what follows, we will consider a benchmark scenario with $Q^2_{\max}\leq 1\,\text{GeV}^2$, which leads to an upper boundary $E\lesssim 3\,\text{GeV}$, since only the total cross section will be involved in both estimating the event rate and constraining the Wilson coefficients. Eventually, we find a proper beam with an intensity up to $150\,\mu$A and a beam energy ranging from $1.1$ to $4.5\,\text{GeV}$, which has been used in the APEX experiment at Jefferson Laboratory (JLab) for hunting sub-GeV dark vector bosons~\cite{Abrahamyan:2011gv,Essig:2010xa}. To maximize the event rate of the scattering process in our benchmark scenario, we will set the beam energy at $3\,\text{GeV}$ and keep the beam intensity at $150\,\mu$A. Nevertheless, it is important to point out that an electron beam with higher beam energy and intensity, as will be shown in Fig.~\ref{fig:Evary}, is certainly favored, but the analyses have to be conducted within the proper $Q^2$ range established above.     

For the proton targets, we favor a liquid hydrogen target with density $\rho=71.3\times 10^{-3}\text{g}/\text{cm}^3$ at $20\,\text{K}$ and $207$--$228\,\text{kPa}$. Such a target has been utilized in the Qweak experiment at JLab with an electron beam of $E=1.16\,\text{GeV}$ and $I=180\,\mu\text{A}$~\cite{Allison:2014tpu}. It is well-known that the liquid target length is limited due to the heating problem and, to break the limit, a cooling system is necessary. Given that the energy (or heat) $H$ stored in the target as an electron beam passes through is given by $H=L \rho dE/dL$~\cite{Zyla:2020zbs},
where $dE/dL$ represents the mean rate of the electron energy loss in units of $\text{MeV}\,\text{g}^{-1}\,\text{cm}^2$, and the cooling power $P$ is defined as $P=H I$,  
with a $3$-$\text{kW}$ cooling system, the Qweak experiment sets the length of its liquid hydrogen target to be around $35\,\text{cm}$~\cite{Allison:2014tpu}.

Assuming that the same cooling system can be applied to our case, we find that the maximal length of the liquid hydrogen target is about $40\,\text{cm}$. It should be noted that more ambitious targets for future experiments have been proposed. For example, the P2 target at Mainz will be $60\,\text{cm}$ long and able to absorb $4$-$\text{kW}$ heat for a $150$-$\mu\text{A}$ electron intensity~\cite{Becker:2018ggl}, and the $150$-$\text{cm}$-long JLab M\o ller target will be capable of absorbing $5\,\text{kW}$ at $85\,\mu\text{A}$~\cite{Benesch:2014bas}. Finally, based on the aforementioned setup, we summarize in Table~\ref{table:parameters} our preferred experimental parameters. 

\begin{table*}[ht]
	\renewcommand\arraystretch{1.3} 
	\tabcolsep=0.35cm
	\centering
	\begin{tabular}{|c|c|c|c|c|}
		\hline \hline
		\multicolumn{2}{|c|}{Electron beam}&
		\multicolumn{2}{c|}{Liquid hydrogen target}& Luminosity\\
		\cline{1-4}
		Energy~(GeV)& Intensity~($\mu$A) & Length~(cm)& Density~(g/$\text{cm}^3$)& ($\text{cm}^{-2}\text{s}^{-1}$)\\
		\hline
		3 & 150 & 40 & $71.3\times 10^{-3}$ & $1.6\times 10^{39}$\\ 
		\hline \hline
	\end{tabular}
	\caption{Summary of the experimental parameters for the low-energy scattering experiments.} 
	\label{table:parameters} 
\end{table*}

\subsection{Competitive and complementary}

Based on the experimental setup in Table~\ref{table:parameters} and for 
a running time of 1 year (yr), the sensitivity to the effective WCs of the low-energy $\mathcal{L}_{\text{eff}}$ is given by 
\begin{align}
	&\frac{1.4\pi^2}{G^2_F\alpha^2_e}\Big\{20\left(|g^{LL}_{V}|^2+|g^{RR}_{V}|^2\right)
	+8.4\left(|g^{RL}_V|^2+|g^{LR}_V|^2\right) \nonumber \\[0.15cm] 
	&\quad +0.30\,\text{Re}\left[g^{RL}_Vg^{RR*}_V+g^{LL}_{V}g^{LR*}_{V}\right] \nonumber \\[0.15cm]
	&\quad +4.2\left(|g^R_S|^2+|g^L_S|^2\right)+124\left(|g^R_T|^2+|g^L_{T}|^2\right)\nonumber\\[0.15cm]
	&\quad -26\,\text{Re}\left[g^R_Sg^{R*}_T+g^L_Sg^{L*}_T\right]\Big\}\nonumber \\[0.15cm]
	&\quad \!\lesssim\! \left(\frac{N\, \text{events}}{1\, \text{event}}\right)\!\left( \frac{1\, \text{yr}}{t}\right)
	\!\left(\frac{2.1\times10^{22}\, \text{cm}^{-3}}{\rho_T}\right)\!\left(\frac{40\, \text{cm}}{L}\right) \nonumber \\[0.15cm] 
	&\quad \quad\!\times \!\left(\frac{9.4\times10^{14}\, \text{s}^{-1}}{\phi}\right)\!\left(\frac{100\%}{\epsilon_{\Lambda_c}}\right)\!\left(\frac{100\%}{\epsilon_e}\right),\label{eq:constrain_general}
\end{align}
where $\epsilon_{\Lambda_c}$ and $\epsilon_e$ denote the detecting efficiency of the generated $\Lambda_c$ and the scattered electron, respectively.\footnote{Note that it might be unnecessary to detect both the produced $\Lambda_c$ and the scattered electron.}  
As done in Sec.~\ref{subsection:LFC_other_constraints}, once the interference terms are neglected, we can obtain constraints on $|g_{V}|^2 $, $|g_S|^2$, and $|g_T|^2$ from Eq.~\eqref{eq:constrain_general}, which are also collected in Table~\ref{table:constraints_LFC}. It can be seen that, compared with other processes except the leptonic $D$-meson decay, significant improvements in constraining the effective WCs can be made through the low-energy scattering process. Meanwhile, this indicates that the scattering process can provide a further complementarity to the charmed-hadron weak decays and the high-$p_T$ dilepton invariant mass tails.  

It should be mentioned that our results can be weakened by the non-$100\%$ detecting efficiency of the produced particles. Especially for the $\Lambda_c$ baryon, it is hard to keep track of all its decay products. In practice, one may focus only on one of its decay channels, such as $\Lambda_c\to p K^-\pi^+$ with its decay fraction of about $6.28\%$~\cite{Zyla:2020zbs}. However, even so, the scattering process can still provide competitive results than from the semileptonic $D$-meson decay and the high-$p_T$ dilepton invariant mass tails.  

\subsection{\boldmath $e^-p\to e^-\Lambda_c$ in the survived LQ models}

Now we are ready to perform an event-rate estimation for the scattering process in the survived LQ models. The total cross section in the scalar $R_2$ model is given by
\begin{align}
	\sigma_{R_2}=&\left[8.4\left(|g^{RL}_V|^2+|g^{LR}_V|^2\right)
	+2.8\left(|g^R_S|^2+|g^L_{S}|^2\right)\right]\nonumber \\[0.15cm]
	&\times 10^{-4}\, \text{GeV}^2,\label{eq:total_cross_R2}
\end{align}
where the RG running effects in Eq.~\eqref{eq:RG_R2} have been taken into account. Following the same procedure, we have also computed the total cross section in the vector $U_{3}$ ($\tilde{U}_1$) model, with the final result given by
\begin{align}
	\sigma_{U_{3}(\tilde{U}_1)}&=20\,|g^{LL(RR)}_{V}|^2\times 10^{-4}\,\text{GeV}^2. \label{eq:total_cross_V}
\end{align}

We will assume that only one non-vanishing WC contributes to the process at a time, and take the upper limits in Eqs.~\eqref{eq:const_x4} and \eqref{eq:const_v23} for each LQ. Then, supposing a running time of 1 yr, we evaluate the expected event rates in units of number per year ($N$/yr) in different LQ models. The final results are given in Table~\ref{table:event_forcast_LFC}. It can be seen that, if the contributions from scalar and tensor operators dominate, it would be less promising to observe the LFC scattering process, mainly because of the stringent constraints on these operators from the leptonic $D$-meson decay. In addition, the vector LQ models are expected to generate more events than the scalar one. 

\begin{table}[ht]
	\renewcommand\arraystretch{1.3} 
	\tabcolsep=0.65cm
	\centering
	\begin{tabular}[t]{|c|c|c|}
		\hline \hline
		Models &$g^i_V$ & $g^{L,R}_S$  \\ \hline
		$R_2$ & 43 & 0.25\\ 
		$U_3$ & 103 & $\backslash$ \\ 
		$\tilde{U}_1$ & 103 & $\backslash$ \\ 
		\hline \hline
	\end{tabular} 
	\caption{Summary of the event-rate estimations for $e^-p\rightarrow e^-\Lambda_c$ in the three survived LQ models, where $i=LR,RL$ and $LL$~($RR$) for $R_2$ and $U_{3}$~($\tilde{U}_1$), respectively. Note that only one non-vanishing WC is assumed to saturate the process at a time, and the event rate is given in units of $N$/yr. The entries with ``$\backslash$'' mean that no estimations are available due to the absence of the corresponding WCs in the LQ models. } 
	\label{table:event_forcast_LFC} 
\end{table} 

\begin{figure}[ht]
	\centering
	\includegraphics[width=0.48\textwidth]{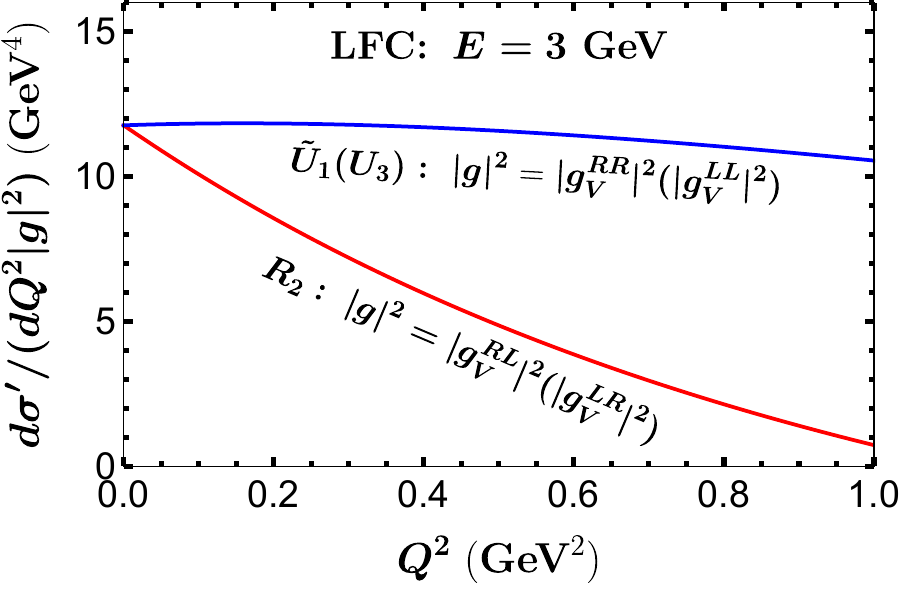}
	\caption{Differential cross section $d\sigma'/(dQ^2|g|^2)$, with $\sigma' =(256\pi m^2_p E^2)\sigma$, in the $R_2$ (red curve) and $U_3$~($\tilde{U}_1$) (blue curve) models. Note that we have neglected the scalar and tensor contributions in the $R_2$ model, because of the stringent constraints from the leptonic $D$-meson decay (see Table~\ref{table:constraints_LFC}).}
	\label{fig:SigmaQ2}  
\end{figure} 

Rather than the total cross section, one may be more interested in how the differential cross sections vary in the available kinematic region for the three different LQ models. As shown in Fig.~\ref{fig:SigmaQ2}, after factoring out $|g|^2$, the differential cross section remains roughly constant in the $U_3$ and $\tilde{U}_1$ models, whereas it decreases rapidly as $Q^2$ approaches $Q^2_{\max}$ in the $R_2$ model. This is because the operators $j^L_VJ^L_V$, $j^R_VJ^R_V$ and $j^R_VJ^L_V$, $j^L_VJ^R_V$ (see their definitions in Appendix~\ref{appendix:cross section}) contribute differently to the differential cross section. As can also be inferred from Eqs.~\eqref{eq:Dee}, \eqref{eq:DPee}, \eqref{eq:qqee} and~\eqref{eq:constrain_general}, such a distinct phenomenon is the unique feature of low-energy scattering process. This suggests that, in contrast to the vector LQs $U_3$ and $\tilde{U}_1$, searching for the scalar LQ $R_2$ is preferred to be conducted in the low-$Q^2$ region. Furthermore, one may exploit the unique feature to distinguish the scalar LQ from the vector ones in future low-energy scattering experiments. For example, if indisputable signals are observed in both $D^+\to \pi^+ e^+e^-$ decay and $e^-p\to e^-\Lambda_c$ scattering, an analysis of $d\Gamma(D^+\to \pi^+ e^+e^-)/d\sigma(e^-p\to e^-\Lambda_c)$ in their ``common'' kinematic region, say $Q^2\in [0.04,0.9]\,\text{GeV}^2$, would indicate the presence of a scalar or a vector LQ.  
  
\begin{figure}[ht]
	\centering
	\includegraphics[width=0.48\textwidth]{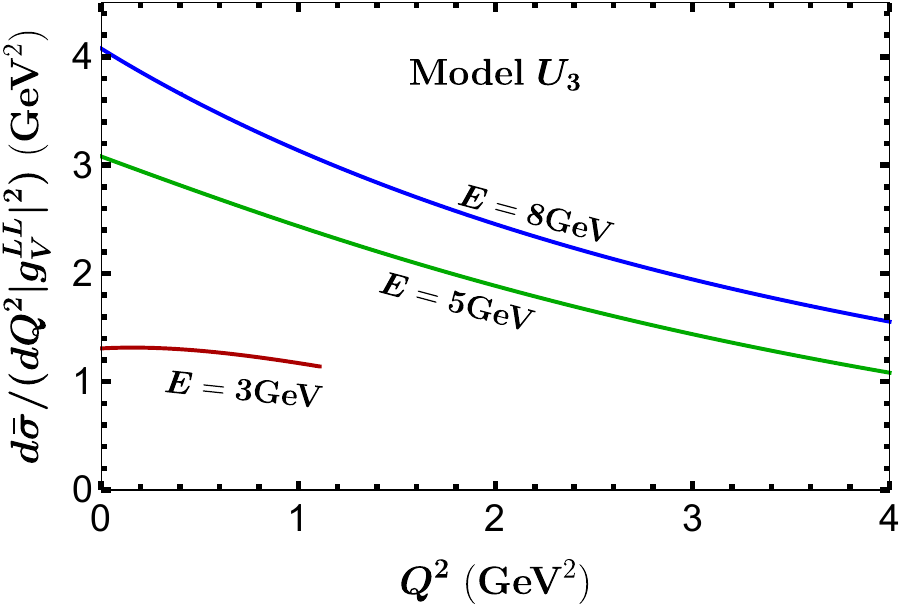}
	\caption{Differential cross section $d\bar{\sigma}/(dQ^2|g^{LL}_V|^2)$ in the LQ model $U_3$, with different electron beam energies, where $\bar{\sigma}$ is defined as $\bar{\sigma} =(256\pi m^2_p)\sigma=\sigma'/E^2$. }
	\label{fig:Evary} 
\end{figure} 

We now explore the dependence of the differential cross section on the electron beam energy $E$. Here we take the LQ model $U_3$ as an example. As shown in Fig.~\ref{fig:Evary}, along with the increase of $E$, the available kinematic region of $Q^2$ expands considerably. Meanwhile, due to the enlarged $Q^2$ region, it becomes clear that the differential cross section that remains constant in the $E=3\,\text{GeV}$ case falls gradually---but still not as dramatically as in the $R_2$ model. Such a trend becomes more prominent for a higher $E$. Finally, it can be seen that a high beam energy clearly favors a high event rate. 
 
Before concluding this section, let us make the following interesting comment. Given that, besides the vector operators $j^L_VJ^R_V$ and $j^R_VJ^L_V$, both the scalar ($j^L_SJ^L_S$ and $j^R_SJ^R_S$) as well as the tensor ($j^L_TJ^L_T$ and $j^R_TJ^R_T$) operators can also contribute to the scattering process in the $R_2$ model, it is tempting to ask if their contributions can be distinguished from those of the vector interaction through the analysis of the differential cross section discussed above. Unfortunately, this cannot be achieved at the moment, because, on the one hand, it is difficult to separate these contributions via the differential cross section and, on the other hand, the scalar contribution is severely constrained by the leptonic decay $D^0\to e^+ e^-$. However, if the constraints on both the vector and scalar (tensor) contributions can reach a similar level in the future---so that both contributions can equally matter---it would be possible to fully distinguish the contributions from the operators $j^L_VJ^R_V$, $j^R_VJ^L_V$, $j^L_SJ^L_S$ ($j^L_TJ^L_T$), and $j^R_SJ^R_S$ ($j^R_TJ^R_T$) present in the $R_2$ model through the low-energy polarized scattering process $e^-p\to e^-\Lambda_c$~\cite{Yan-2022}.


\section{\boldmath LFV FCNC process $e^-p\to \mu^-\Lambda_c$}
\label{sec:LFV_FCNC}

We now turn to discuss the LFV FCNC scattering process $e^-p\to \mu^-\Lambda_c$. 

\subsection{Kinematics and beam energy selection}

The total cross section of the scattering process $e^-(k)+p(P)\to \mu^-(k')+\Lambda_c(P')$ 
can be calculated in the same way as for the LFC case. But its spin-averaged amplitude squared $\overline{|\mathcal{M}|}^2$ contains some additional terms induced by the non-negligible muon mass $m_{\mu}$.

Due to the presence of $m_{\mu}$, the kinematics and the lower bound on the electron beam energy $E$ of the LFV scattering process are also changed. We find that the variable $q^2$ must now satisfy the condition
\begin{align}
	\frac{\alpha-E\sqrt{\lambda_{\mu}}}{m_p+2E} \leq
	q^2 \leq \frac{\alpha +E\sqrt{\lambda_{\mu}}}{m_p+2E},\label{eq:LFV_q^2}
\end{align}
with
\begin{align}
	\alpha &\!\equiv \! E(m_{\Lambda_c}^2\!-\! m_p^2\!+\!m^2_{\mu}\!-\! 2m_pE)\!+\!m_p m^2_{\mu}, \nonumber \\[0.2cm]
	\lambda_{\mu} &\!\equiv\! m_{\Lambda_c}^4\!+\!(m_p^2\!+\!2m_pE\!-\!m^2_{\mu})^2\!-\!2m_{\Lambda_c}^2(m_p^2\!+\!2m_pE\!+\!m^2_{\mu}). \nonumber 
\end{align}
This in turn indicates the beam energy condition
\begin{align}
	E\geq \frac{(m_{\Lambda_c}+m_{\mu})^2-m_p^2}{2m_p}.\label{eq:LFV_E_cond}
\end{align}
Compared with Eq.~\eqref{eq:LFC_E_cond}, Eq.~\eqref{eq:LFV_E_cond} produces a slightly higher $E_{\min}$, rendering a tiny difference of $Q^2_{\max}$ in the low-$E$ region. However, due to the small ratios $m_{\mu}/m_{\Lambda_c}$ and $m_{\mu}/E$, such a difference can be safely neglected. For our benchmark scenario with $Q^2_{\max}\leq 1\,\text{GeV}^2$, we find that the $3$-$\text{GeV}$ electron beam selected in the LFC case is also suitable for the LFV scattering process. Thus, interestingly, one can explore both the LFC and LFV scattering processes at the same time with the same experimental setup. 

\subsection{Relevant experimental constraints}

We give a brief survey of the best relevant constraints on $[g]^{\mu e,cu}$ from the LFV charmed-hadron weak decays and the high-$p_T$ dilepton mass tails in $pp(q\bar{q})\to e^{\pm}\mu^{\mp}$.

\begin{itemize}
	\item \textit{Constraint from $D^0\to e^{-}\mu^{+}$} 
\end{itemize}

We start with the LFV leptonic decay of the neutral $D$ meson. The branching fraction of $D^0\to e^{-}\mu^{+}$ decay can be found, \textit{e.g.}, in Refs.~\cite{deBoer:2015boa,Bause:2019vpr}. Using the experimental limit on the branching fraction, $\mathcal{B}(D^0\to e^{-}\mu^{+})<1.3\times 10^{-8}$ at $90\%$ C.L.~\cite{Aaij:2015qmj}, we obtain 
\begin{align}
	&|K_S-K'_S- 0.04(K_9-K'_9)|^2 \nonumber \\[0.15cm]
	&\quad +|K_P-K'_P+0.04(K_{10}-K'_{10})|^2\lesssim 0.01,
\end{align}
which is consistent with the latest result presented in Ref.~\cite{Bause:2019vpr}. With the relations defined in Eq.~\eqref{eq:relation}, we can rewrite the constraint as 
\begin{align}
	\frac{\pi^2}{G^2_F\alpha^2_e}&\bigg\{0.08\text{Re}\left[g^L_S\left(g^{RL}_V-g^{RR}_V\right)^*
	-g^R_S\left(g^{RR}_V-g^{LR}_V\right)^*\right]\nonumber \\[0.15cm]
	&+|g^R_S|^2+|g^L_S|^2\bigg\}\lesssim 0.01,
\end{align}
where we have neglected the vector-vector interference terms, due to the tiny coefficients associated with them. 

\begin{itemize}
	\item \textit{Constraint from $D^+\to \pi^+ e^- \mu^+$}
\end{itemize}

Theoretical studies of the LFV semileptonic $D^+\to \pi^+ e^- \mu^+$ decay have been conducted in Refs.~\cite{deBoer:2015boa,Bause:2019vpr}. Considering the experimental limit, $\mathcal{B}(D^+\to \pi^+ e^- \mu^+) < 3.6\times 10^{-6}$ at $90\%$ C.L.~\cite{Lees:2011hb}, we obtain the constraint
\begin{align}
	& 2.5\left(|K_{9}+K'_{9}|^2+|K_{10}+K'_{10}|^2\right) +5.2\left(|K_S+K'_S|^2 \right. \nonumber \\[0.15cm] 
	&\left. \quad +|K_P+K'_P|^2\right) +0.72\left(|K_T|^2+|K_{T5}|^2\right) \nonumber \\[0.15cm] 
	&\quad +0.63\,\text{Re}\left[\left(K_{9}+K'_{9}\right)K^*_T- \left(K_{10}+K'_{10}\right)K^*_{T5}\right] \nonumber \\[0.15cm] 
	&\quad +1.0\,\text{Re}\left[\left(K_{10}+K'_{10}\right) \left(K^*_{P}+K^{\prime\ast}_{P}\right)\right. \nonumber \\[0.15cm] 
	&\quad \left. -\left(K_{9}+K'_{9}\right)\left(K^*_{S}+K^{\prime\ast}_{S}\right)\right] \lesssim 100,
\end{align}
which is consistent with the corrected result presented in Ref.~\cite{Bause:2019vpr}. Rewriting them in terms of $g$, we get  
\begin{align}
	\frac{\pi^2}{G^2_F\alpha^2_e}&\Big\{5.2\,(|g^L_S|^2+|g^R_S|^2)+2.9\,(|g^L_T|^2+|g^R_T|^2)\nonumber \\[0.15cm]
	&+2.5\,\big(|g^{RL}_V|^2+|g^{LR}_V|^2+|g^{RR}_V|^2+|g^{LL}_V|^2\big)\nonumber \\[0.15cm]
	&+2.0\,\text{Re}[g^{RL}_Vg^{RR*}_V+g^{LL}_Vg^{LR*}_V]\big)-1.0\,\text{Re}[g^{LL}_{V}g^{R*}_{S}]\nonumber \\[0.15cm]
	&+1.2\,\text{Re}[g^{RL}_Vg^{L*}_{T}+g^{LL}_Vg^{R*}_{T}]\Big\}\lesssim 100.
\end{align}

\begin{itemize}
	\item \textit{Constraints from high-$p_T$ dilepton mass tails}
\end{itemize}

Constraints on the effective WCs from the analyses of high-$p_T$ dilepton invariant mass tails in $pp(q\bar{q})\to e^{\pm}\mu^{\mp}$ have recently been worked out in Ref.~\cite{Angelescu:2020uug}. They recast the latest ATLAS analysis with $36.1\,\text{fb}^{-1}$ of $13$-$\text{TeV}$ data~\cite{Aaboud:2018jff}, and obtained the following bounds at $90\%$ C.L.:
\begin{align}
	|g^i_V|\!\lesssim \! 1.1\frac{G_F\alpha_e}{\pi},\, 
	|g^{L,R}_S|\! \lesssim \! 2.4\frac{G_F\alpha_e}{\pi},\,  
	|g^{L,R}_T|\!\lesssim\!  0.44\frac{G_F\alpha_e}{\pi},
\end{align} 
where $i=LL,RR,LR,RL$. Note that the RG running effects neglected in Ref.~\cite{Angelescu:2020uug} have been taken into account. Being the same as observed in the $pp(q\bar{q})\to e^+e^-$ case, no sensible constraints can be set on the interference terms as well.

\begin{table}[ht]
	\renewcommand\arraystretch{1.3} 
	\tabcolsep=0.08cm
	\centering
	\begin{tabular}[t]{|c|c|c|c|c|}
		\hline \hline
		Processes &$\big|g^{LL,RR}_V\big|^2$&$\big|g^{LR,RL}_V\big|^2$ & $\big|g^{L,R}_S\big|^2$& $\big|g^{L,R}_T\big|^2$  \\ \hline
		$D^0\to e^-\mu^+$~\cite{Aaij:2015qmj} & $\backslash$& $\backslash$ & 0.010& $\backslash$ \\
		$D^+\to \pi^+ e^-\mu^+$~\cite{Lees:2011hb} & 40& 40& 19 & 34 \\
		$pp(q\bar{q})\to e^{\mp}\mu^{\pm}$~\cite{Aaboud:2018jff} & 1.2 & 1.2 & 5.8&0.19 \\
		$e^-p\to \mu^-\Lambda_c$ & 0.039 & 0.091 & 0.18&0.0063 \\ \hline \hline
	\end{tabular}
	\caption{Constraints on the WCs $[g]^{\mu e,cu}$ at $90\%$ C.L. from the LFV (semi)leptonic $D$-meson decays, the high-$p_T$ dilepton invariant mass tails, and the $e^-p\to \mu^-\Lambda_c$ scattering process in the framework of a general low-energy $\mathcal{L}_{\text{eff}}$ specified by Eq.~\eqref{eq:Lag_LQ}. Note that the common factor $G^2_F\alpha^2_e/\pi^2$ has been factored out. The entries with ``$\backslash$'' mean that the processes in the first column put no constraints on the corresponding WCs. } 
	\label{table:constraints_LFV} 
\end{table} 

We summarize in Table~\ref{table:constraints_LFV} the relevant constraints on the WCs $[g]^{\mu e,cu}$. It can be seen that the most stringent constraint on $g_S$ comes from the measurement of the LFV leptonic $D$-meson decay, which clearly constrains neither $g_V$ nor $g_T$. Compared with the LFV semileptonic $D$-meson decay, on the other hand, the analysis of the high-$p_T$ dilepton invariant mass tails in $pp(q\bar{q})\to e^{\mp} \mu^{\pm}$ processes can set more severe limits on $g_V$ and $g_T$. 

Constraints on $[g]^{\mu e,cu}$ in specific LQ models can be read out from Table~\ref{table:constraints_LFV} straightforwardly. For the scalar $R_2$ model, even with the RG running effects (see Eq.~\eqref{eq:RG_R2}) taken into account, the constraint on $g_S$ is still dictated by the LFV leptonic $D$-meson decay, whereas the boundary of $g_V$ is determined by $pp(q\bar{q})\to e^{\mp}\mu^{\pm}$ processes. We thus obtain the following constraints for the $R_2$ model:
\begin{align}
	&|g^{RL,LR}_V|^2\lesssim 1.2 \left(\frac{G^2_F\alpha_e^2}{\pi^2}\right),\nonumber \\
	&|g^{L,R}_S|^2\simeq 88 |g^{L,R}_T|^2 \lesssim 0.010 \left(\frac{G^2_F\alpha_e^2}{\pi^2}\right).\label{eq:const_LFV_R2}
\end{align}
For the vector $U_{3}$~($\tilde{U}_1$) model, we find
\begin{align}
	|g^{LL(RR)}_V|^2\lesssim 1.2 \left(\frac{G^2_F\alpha_e^2}{\pi^2}\right). \label{eq:const_LFV_v23}
\end{align} 

\subsection{Competitive and complementary}

Utilizing the same experimental setup, we can set limits on the WCs through the LFV scattering process. In the context of low-energy $\mathcal{L}_{\text{eff}}$, we make the same assumptions as in the LFC case, and obtain the sensitivity 
\begin{align}
	&\frac{1.4\pi^2}{G^2_F\alpha^2_e}\Big\{18\left(|g^{LL}_{V}|^2\!+\!|g^{RR}_{V}|^2\right)\!
	+\!7.7\left(|g^{RL}_V|^2\!+\!|g^{LR}_V|^2\right)\nonumber \\[0.15cm] 
	&\quad \!+\!0.28\,\text{Re}\!\left[g^{RL}_Vg^{RR*}_V\!+\!g^{LL}_{V}g^{LR*}_{V}\right]\!+\!3.9\left(|g^R_S|^2\!+\!|g^L_S|^2\right)\!\nonumber \\[0.15cm] 
	&\quad \! +\!4.4\,\text{Re}[g^R_Sg_V^{RR*}\!+\!g^{L}_Sg^{LL*}_V]\!+\!0.63\,\text{Re}[g^R_Sg_V^{RL*}\!+\!g^{L}_Sg^{LR*}_V]\nonumber \\[0.15cm] 
	&\quad \!-\!33\,\text{Re}[g^{R}_Tg^{RR*}_{V}\!+\!g^{L}_Tg^{LL*}_{V}]\!+\!3.0\,\text{Re}[g^{R}_Tg^{RL*}_{V}\!+\!g^{L}_Tg^{LR*}_{V}]\nonumber \\[0.15cm] 
	&\quad\! +\!112\left(|g^R_T|^2\!+\!|g^L_{T}|^2\right)\!-\! 23\,\text{Re}\!\left[g^R_Sg^{R*}_T\!+\!g^L_Sg^{L*}_T\right]\Big\}\nonumber \\[0.15cm] 
	&\quad \!\lesssim \!\left(\frac{N\,\text{events}}{1\,\text{event}}\right)\! \left(\frac{2.1\times10^{22}\,\text{cm}^{-3}}{\rho_T}\right)\!\left(\frac{40\ \text{cm}}{L}\right)\!\left( \frac{1\ \text{yr}}{t}\right)\nonumber \\[0.15cm] 
	&\quad \quad\!\times\! \left(\frac{9.4\times10^{14}\ \text{s}^{-1}}{\phi}\right)\!\left(\frac{100\%}{\epsilon_{\Lambda_c}}\right)\!\left(\frac{100\%}{\epsilon_{\mu}}\right),\label{eq:constrain_general_LFV}
\end{align}
where $\epsilon_{\mu}$ denotes the detecting efficiency of the muon lepton generated in the LFV scattering process.

Focusing on the constraints on $|g_{V}|^2$, $|g_S|^2$, and $|g_T|^2$, we compare in Table~\ref{table:constraints_LFV} our results with those obtained from the $D$-meson weak decays and the high-$p_T$ dilepton invariant mass tails. It can be seen that appreciable improvement on the constraints can be made through the LFV scattering experiment. Even the obtained limit of $|g_S|^2$, though not as stringent as that set by the LFV leptonic $D$-meson decay, is still very competitive.

As in the LFC case, our constraints can be weakened by the imperfect detecting efficiency of the $\Lambda_c$ baryon, $\epsilon_{\Lambda_c}$. However, even if the produced $\Lambda_c$ is solely detected via the decay $\Lambda_c\to pK^-\pi^+$, our results are still comparable to those from the high-$p_T$ dilepton invariant mass tails.   

For both the LFC and LFV scattering experiments, further improvement could be made by considering an electron beam with a higher intensity, if available in the future, and a hydrogen gas target working in more severe conditions. Note that a gas target is free from the boiling problem, and consequently its luminosity is not limited by the cooling power. However, higher intensity or longer target length would be preferred to compensate its relatively lower number density. 

\subsection{\boldmath $e^-p\to \mu^-\Lambda_c$ in the survived LQ models}

We now evaluate the total cross sections specifically in the three LQ models. For the scalar $R_2$ model, we obtain
\begin{align}
	\sigma_{R_2} &= \Big[7.7\left(|g^{RL}_V|^2+|g^{LR}_V|^2\right)
	+2.7\left(|g^R_S|^2+|g^L_{S}|^2\right)\nonumber \\[0.15cm]
	&+0.95\,\text{Re}[g^R_Sg_V^{RL*}+g^{L}_Sg^{LR*}_V ]\Big]\times 10^{-4}\,\text{GeV}^2,\label{eq:total_cross_LFV_R2}
\end{align}
where the RG running effects in Eq.~\eqref{eq:RG_R2} have been considered. Note that the scalar-vector interference terms emerge because of the non-vanishing $m_{\mu}$. Similarly, the total cross section in the vector $U_{3}$ ($\tilde{U}_1$) model is computed to be  
\begin{align}
	\sigma_{U_{3}(\tilde{U}_1)}&=18\,|g^{LL(RR)}_{V}|^2\times 10^{-4}\,\text{GeV}^2. \label{eq:total_cross_LFV_V}
\end{align} 

Let us give a simple event-rate estimation for the scattering experiment in the three different LQ models. As done in the LFC case, we ignore the contributions from the interference terms, take the upper limits given in Eqs.~\eqref{eq:const_LFV_R2} and \eqref{eq:const_LFV_v23} for each LQ model, and always assume a running time of 1 yr. It can be seen from Table~\ref{table:event_forcast_LFV} that, in comparison with the LFC case, fewer events are now expected for the LFV scattering process, mostly because of the relatively more severe constraints on the vector operators. 

\begin{table}[ht]
	\renewcommand\arraystretch{1.3} 
	\tabcolsep=0.60cm
	\centering
	\begin{tabular}[t]{|c|c|c|}
		\hline \hline
		Models &$g^i_V$ & $g^{L,R}_S$  \\ \hline
		$R_2$ & 13 & 0.039 \\
		$U_3$ & 31& $\backslash$ \\
		$\tilde{U}_1$ & 31 & $\backslash$ \\
		\hline \hline
	\end{tabular} 
	\caption{Summary of the event-rate estimations for $e^-p\rightarrow \mu^-\Lambda_c$ in the three survived LQ models, where $i=LR,RL$ and $LL$~($RR$) for $R_2$ and $U_{3}$~($\tilde{U}_1$), respectively. The entries with ``$\backslash$'' mean that no estimations are available due to the absence of the corresponding WCs in the LQ models. }
	\label{table:event_forcast_LFV}
\end{table} 

\begin{figure}[ht]
	\centering
	\includegraphics[width=0.48\textwidth]{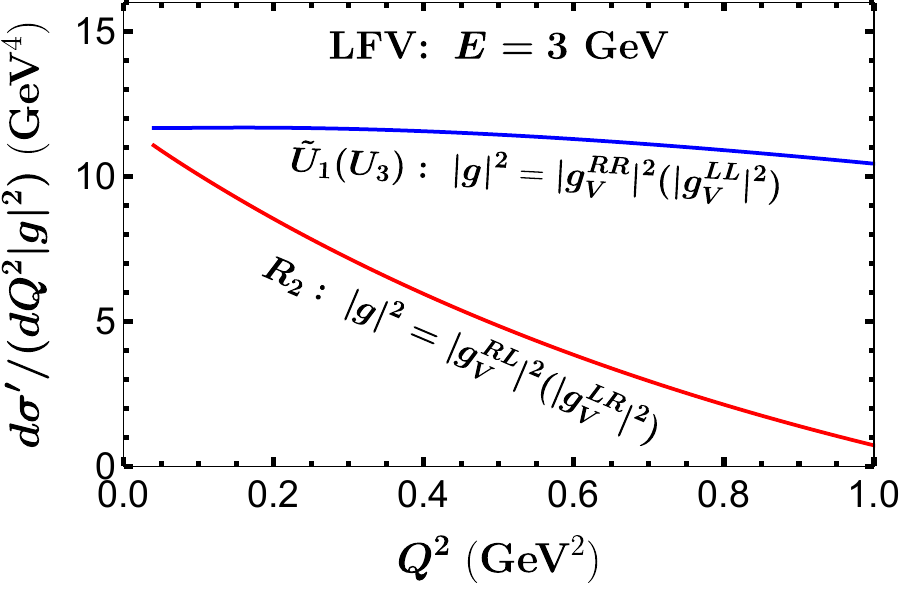}
	\caption{Differential cross section $d\sigma'/(dQ^2|g|^2)$ in the LFV case. The other captions are the same as in Fig.~\ref{fig:SigmaQ2}}
	\label{fig:SigmaQ2_LFV}
	\vspace*{-.1in} 
\end{figure} 

We have also looked into the differential cross sections in the allowed kinematic region. 
As shown in Fig.~\ref{fig:SigmaQ2_LFV}, they behave in the same way as in Fig.~\ref{fig:SigmaQ2}. This is not surprising, since these two scattering processes are identical in respect of kinematics, except the mass difference between the electron and the muon. Although a non-zero $m_{\mu}$ can indeed induce non-vanishing interference terms, such as the last term in Eq.~\eqref{eq:total_cross_LFV_R2} for the $R_2$ model, as well as additional non-interference contributions, they are too small to make a numerical difference.
  
\section{Conclusion}
\label{sec:con}

We have investigated the potential for discovering NP in the charm sector through the low-energy scattering processes $e^-p\to e^-\Lambda_c$ and $e^-p\to \mu^-\Lambda_c$. Focusing on the fixed-target $ep$ scattering experiments, we have studied the kinematics of both scattering processes. It has been shown that, even though the corresponding minimal beam energies vary due to their different kinematics, they can be simultaneously detected with the same experimental setup. Taking account of all the constraints on the beam energies, we have eventually proposed an electron beam with energy of $3$~GeV. Meanwhile, to maximize the chance of observing their signals, we have chosen a liquid hydrogen target with a powerful cooling system. It is intriguing to note that both the beam and the target have already been utilized in different experiments.  
 
Based on the selected experimental setup, we have demonstrated in a model-independent way
that, compared with the charmed-hadron weak decays and the high-$p_T$ dilepton invariant mass tails, the low-energy scattering processes can provide more competitive constraints and, at the same time, build a further complementarity with the charmed-hadron weak decays. On the other hand, in the specific LQ models, we have shown that promising event rates can be expected for both the LFC and LFV scattering processes, with the most stringent constraints on the corresponding WCs from other processes taken as input. 

We have also analyzed the differential cross sections in the allowed kinematic region, and observed that they all decrease gradually as $Q^2$ approaches to its maximum $Q^2_{\max}$. Interestingly, such a trend becomes more distinct for a higher beam energy. Furthermore, we have found that the decreasing rate in the scalar $R_2$ model is always more dramatic than those in the vector $U_{3}$ and $\tilde{U}_{1}$ models, which provides a potential way to distinguish the scalar and vector LQs in future experiments.    

Finally, we would like to remark that, since the majority of our analyses are based on the general low-energy effective Lagrangian, it is straightforward to generalize them into other specific NP scenarios~\cite{Yan-2022}. 

\section*{Acknowledgments}

This work is supported by the National Natural Science Foundation of China under Grant 
Nos.~12135006, 12075097, 12047527, 11675061, and 11775092.

\appendix
\renewcommand{\theequation}{A.\arabic{equation}}
\section{\boldmath Definitions and parametrization of the $\Lambda_c\to p$ form factors}
\label{appendix:form factor}

The form factors for $\Lambda_c\to p$ transition can be conveniently parametrized in 
the helicity basis~\cite{Feldmann:2011xf,Meinel:2017ggx,Das:2018sms}. For the vector and axial-vector currents, their hadronic matrix elements are defined, respectively, by 
\begin{align}
&\langle N^{+}(p,s)|\bar{u}\gamma^{\mu}c|\Lambda_c(p',s')\rangle \nonumber \\[0.15cm]
&\ \ \!=\!\bar{u}_N(p,s)\bigg[f_0(q^2)(m_{\Lambda_c}\!-\!m_N)\frac{q^{\mu}}{q^2}\nonumber \\[0.15cm]
&\ \  \ \ \ \!+\!f_+(q^2)\frac{m_{\Lambda_c}\!+\!m_N}{s_+}
\Big(p'^{\mu}\!+\!p^{\mu}\!-\!(m^2_{\Lambda_c}\!-\!m^2_N)\frac{q^{\mu}}{q^2}\Big) \nonumber \\[0.15cm]
&\ \ \ \ \  \!+\! f_{\perp}(q^2)\Big(\gamma^{\mu}\!-\!\frac{2m_N}{s_+}p'^{\mu}\!-\!\frac{2m_{\Lambda_c}}{s_+}p^{\mu}\Big)\bigg]u_{\Lambda_c}(p',s'), \label{eq:vect}
\end{align}
and 
\begin{align}
&\langle N^{+}(p,s)|\bar{u}\gamma^{\mu}\gamma^5c|\Lambda_c(p',s')\rangle \nonumber \\[0.15cm]
&\ \ \!=\!-\bar{u}_N(p,s)\gamma^5\bigg[g_0(q^2)(m_{\Lambda_c}\!+\! m_N)\frac{q^{\mu}}{q^2}\nonumber \\[0.15cm]
&\ \ \ \ \ \!+\! g_+(q^2)\frac{m_{\Lambda_c}\!-\!m_N}{s_-}
\Big(p'^{\mu}\!+\!p^{\mu}\!-\!(m^2_{\Lambda_c}\!-\! m^2_N)\frac{q^{\mu}}{q^2}\Big) \nonumber \\[0.15cm]
&\ \ \ \ \ \!+\!g_{\perp}(q^2)\Big(\gamma^{\mu}\!+\!\frac{2m_N}{s_-}p'^{\mu}\!-\!\frac{2m_{\Lambda_c}}{s_-}p^{\mu}\Big)\bigg]u_{\Lambda_c}(p',s'), \label{eq:psvect}
\end{align}
where $q=p'-p$ and $s_{\pm}=(m_{\Lambda_c}\pm m_N)^2-q^2$. Note that we have denoted the proton by $N^+$ instead of $p$ to avoid possible confusion with the proton's momentum. 
From Eqs.~\eqref{eq:vect} and \eqref{eq:psvect}, we can obtain the hadronic matrix elements of the scalar and pseudo-scalar currents through the equation of motion,
\begin{align}
&\langle N^{+}(p,s)|\bar{u}c|\Lambda_c(p',s')\rangle \nonumber \\[0.15cm]
&\ \ =\frac{(m_{\Lambda_c}- m_N)}{m_{c}- m_{u}}f_0(q^2)\bar{u}_N(p,s)u_{\Lambda_c}(p',s') , \\[0.2cm]
&\langle N^{+}(p,s)|\bar{u}\gamma^5	c|\Lambda_c(p',s')\rangle \nonumber \\[0.15cm]
&\ \  =\frac{(m_{\Lambda_c}+m_N)}{m_{c}+m_{u}}g_0(q^2)\bar{u}_N(p,s)\gamma^5 u_{\Lambda_c}(p',s'),
\end{align}
where $m_{u(c)}$ denotes the $u(c)$-quark running mass. Finally, the hadronic matrix element for the tensor current is given by 
\begin{align}
&\langle N^{+}(p,s)|\bar{u}i\sigma_{\mu\nu} c|\Lambda_c(p',s')\rangle \nonumber \\[0.15cm]
&\ \ \!=\!\bar{u}_N(p,s) \bigg[2h_+ \frac{p'_{\mu}p_{\nu}\! -\! p'_{\nu}p_{\mu}}{s_+}\!
+\! h_{\perp}\Big(\frac{m_{\Lambda_c}\! +\! m_N}{q^2}\nonumber \\[0.15cm]
&\ \ \ \ \ \! \times\! (q_{\mu}\gamma_{\nu}\!-\! q_{\nu}\gamma_{\mu})\!-\! 2\left(\frac{1}{q^2}\!
+\!\frac{1}{s_+}\right)(p'_{\mu}p_{\nu}\!-\! p'_{\nu}p_{\mu})\Big)\nonumber \\[0.15cm]
&\ \ \ \ \ \!+\! \tilde{h}_+\Big(i\sigma_{\mu\nu}\!-\!\frac{2}{s_-}[m_{\Lambda_c}(p_{\mu}\gamma_{\nu}\!-\!p_{\nu}\gamma_{\mu})\nonumber \\[0.15cm]
&\ \ \ \ \ \!-\!m_N(p'_{\mu}\gamma_{\nu}\!-\!p'_{\nu}\gamma_{\mu})+\!p'_{\mu}p_{\nu}\!-\!p'_{\nu}p_{\mu}]\Big)\nonumber \\[0.15cm]
&\ \ \ \ \ \!+\tilde{h}_{\perp}\frac{m_{\Lambda_c}-m_N}{q^2s_-}\Big((m_{\Lambda_c}^2-m_N^2-q^2)(\gamma_{\mu}p'_{\nu}-\gamma_{\nu}p'_{\mu})\nonumber \\[0.15cm]
&\ \ \ \ \ \!-(m_{\Lambda_c}^2-m_N^2+q^2)(\gamma_{\mu}p_{\nu}-\gamma_{\nu}p_{\mu})\nonumber \\[0.15cm]
&\ \ \ \ \ \!+\!2(m_{\Lambda_c}-m_N)(p'_{\mu}p_{\nu}-p'_{\nu}p_{\mu})\Big)\bigg]u_{\Lambda_c}(p',s'), \label{eq:tensor}
\end{align}
where $\sigma_{\mu\nu}=i[\gamma_{\mu},\gamma_{\nu}]/2$. These form factors satisfy the endpoint relations
\begin{align}
	& f_+(0)=f_0(0), \quad && g_+(0)=g_0(0), \nonumber \\[0.2cm]
	& g_+(q_{\max}^2)=g_\perp(q_{\max}^2), \quad && \tilde{h}_{+}(q_{\max}^2)=\tilde{h}_{\perp}(q_{\max}^2),
\end{align}
where $q_{\max}^2=(m_{\Lambda_c} - m_N)^2$. As the latest LQCD determinations~\cite{Meinel:2017ggx} of these form factors satisfy the two criteria mentioned in Sec.~\ref{sec:models}, we will adopt the results provided in Ref.~\cite{Meinel:2017ggx} in this work. Explicitly, the parametrization of the form factors takes the form~\cite{Bourrely:2008za,Meinel:2017ggx}
\begin{align}\label{eq:formfactorpara}
f(q^{2})=\frac{1}{1-q^{2} /(m_{\text{pole}}^{f})^{2}} \sum_{n=0}^{n_{\max}} a_{n}^{f}\left[z(q^{2})\right]^{n},
\end{align}
with the expansion variable defined as
\begin{align}
z(q^{2})=\frac{\sqrt{t_{+}-q^{2}}-\sqrt{t_{+}-t_{0}}} {\sqrt{t_{+}-q^{2}}+\sqrt{t_{+}-t_{0}}},
\end{align}
where $t_{+}=(m_{D}+m_{\pi})^{2}$ is set equal to the threshold of $D\pi$ two-particle states, and $t_{0}=q_{\max}^{2}$ determines which value of $q^{2}$ gets mapped to $z=0$. In this way, one maps the complex $q^{2}$ plane, cut along the real axis for $q^{2} \geq t_{+}$, onto the disk $|z|<1$. Furthermore, the lowest poles in Eq.~\eqref{eq:formfactorpara} have been factored out before the $z$ expansion. The quantum numbers and masses of these poles in the different form factors, as well as the values of the $z$-expansion parameters for both the nominal ($n_{\max}=2$) and higher-order ($n_{\max}=3$) fits, together with the full covariance matrices can be found in Ref.~\cite{Meinel:2017ggx}.  

\renewcommand{\theequation}{B.\arabic{equation}}
\section{Cross section and kinematics of the LFC (LFV) scattering process}
\label{appendix:cross section}

In Sec.~\ref{sec:models}, we have established that the effective Lagrangian $\mathcal{L}_{\text{eff}}$ at low energy can mediate both the LFC and LFV scattering processes. Given that each of the four-fermion operators in Eq.~\eqref{eq:Lag_LQ} consists of one leptonic ($j$) and one hadronic ($J$) current, for the benefit of later discussions, we rewrite it as 
$\mathcal{L}_{\text{eff}}=\sum g_{\alpha\beta} j_{\alpha} J_{\beta}$, where
\begin{align}
j^{R,L}_{S}&=\bar{\ell}P_{R,L}\ell, \quad \quad \quad \quad \   J^{R,L}_{S}=\bar{q}P_{R,L} q, \nonumber \\[0.2cm]
(j^{R,L}_{V})^{\mu}&=\bar{\ell}\gamma^{\mu}P_{R,L} \ell,  \quad \quad  (J^{R,L}_{V})^{\mu}= \bar{q}\gamma^{\mu}P_{R,L} q, \nonumber \\[0.2cm]
(j^{R,L}_{T})^{\mu \nu}&= \bar{\ell}\sigma^{\mu \nu} P_{R,L} \ell,\quad   (J^{R,L}_{T})^{\mu \nu}= \bar{q}\sigma^{\mu \nu} P_{R,L} q. \label{eq:operators}
\end{align}
Obviously, for both the scalar and tensor interactions, $(\alpha,\beta)$=$(R,R)$, $(L,L)$, whereas for the vector operators, $(\alpha,\beta)$=$(R,R)$, $(L,L)$, $(L,R)$, and $(R,L)$. 

The differential cross section of the LFC scattering process $e^-(k)+p(P)\to e^-(\mu^-)(k')+\Lambda_c(P')$ is given by 
\begin{align}
d\sigma&\!=\!\frac{1}{F}\frac{d^3\pmb{k}'}{(2\pi)^3}\frac{1}{2E'}\frac{d^3\pmb{p}'}{(2\pi)^3}\frac{1}{2E_{\Lambda_c}}
\overline{|\mathcal{M}|}^2(2\pi)^4\delta^4(P\!+\!k\!-\!P'\!-\! k') \nonumber \\[0.15cm]
&\!=\!\frac{1}{F}\frac{d^3\pmb{k}'}{(2\pi)^2}\frac{1}{2E'}
\overline{|\mathcal{M}|}^2\delta(P'^2-m_{\Lambda_c}^2),\label{eq:diff_cross}
\end{align}
where the flux factor $F=4[(P\cdot k)^2-m^2_em^2_p]^{1/2}$, and we have performed the integration over $\pmb{p}'$ by using the $\delta$ function in the second step above.

Since the electron mass $m_e$ is much smaller than those of other particles and the beam energy $E$, it can be safely neglected. Due to its relatively bigger mass $m_{\mu}$, the muon remains, however, massive and its mass should be kept in the LFV case. With the definition $q \equiv k-k'=P'-P$, the leftover $\delta$ function in Eq.~\eqref{eq:diff_cross} can be simplified as 
\begin{equation}
\delta(P'^2-m^2_{\Lambda_c})=\frac{\delta(\cos\theta-\cos\theta_0)}{2\,E\,E'}, \label{eq:relation_LFC}
\end{equation}
with 
\begin{equation}
\cos\theta_0 =\frac{m_{\Lambda_c}^2-m_p^2-2m_p(E-E')+2EE'}{2\,E\,E'},\label{eq:theta_LFC}
\end{equation}
for the LFC scattering process, and 
\begin{equation}
\delta(P'^2-m^2_{\Lambda_c})=\frac{\delta(\cos\theta-\cos\theta_0)}{2\,E\,|\pmb{k}'|}, \label{eq:relation_LFV}
\end{equation}
with 
\begin{equation}
\cos\theta_0\! =\! \frac{m_{\Lambda_c}^2\!-\!m_p^2\!-\!m_{\mu}^2\!-\!2m_p(E-E')\!+\!2\,E\,E'}{2\,E\,|\pmb{k}'|}, \label{eq:theta_LFV}
\end{equation}
for the LFV scattering process, where $\theta$ is the scattering angle between $\pmb{k}'$ and $\pmb{k}$. One can then use the remaining $\delta$ functions (Eqs.~\eqref{eq:relation_LFC} and \eqref{eq:relation_LFV}) to get rid of the angular integration in Eq.~\eqref{eq:diff_cross}. Meanwhile, from the relation $P'^2=(q+P)^2=m^2_{\Lambda_c}$, it is easy to find that $dE'=dq^2/2m_p$. Now the total cross sections for the two scattering processes can be universally written as
\begin{equation}
\sigma =\frac{1}{64\pi m_{p}^2 E^2}\int^{q^2_{\max}}_{q^2_{\min}}  dq^2\, \overline{|\mathcal{M}|}^2.\label{eq:total_cross}
\end{equation}
The kinematic range of $q^2$ can be determined through the condition $|\cos\theta_0|\leq 1$,   
from which the corresponding $[q^2_{\min}, q^2_{\max}]$ in both cases can be obtained (see Eqs.~\eqref{eq:LFC_Q2_range} and \eqref{eq:LFV_q^2}).   

Finally, in terms of the low-energy effective Lagrangian $\mathcal{L}_{\text{eff}}=\sum g_{\alpha\beta} j_{\alpha} J_{\beta}$, we can write the amplitudes $\mathcal{M}$ for the two scattering processes as 
\begin{equation}
\mathcal{M}=\sum g_{\alpha\beta}\langle \ell^{(')} (k', r')| j_{\alpha}|\ell(k,r)\rangle \langle \Lambda_c(P', s')|J_{\beta}|p(P, s)\rangle, 
\end{equation} 
where $r$ and $s$ ($r'$ and $s'$) denote the spins of the initial (final) lepton and baryon respectively, and the hadronic matrix elements $\langle\Lambda_c(P',s')|J_{\beta}|p(P, s)\rangle$ are given by the complex conjugate of $\langle p(P, s)|J^{\dagger}_{\beta}|\Lambda_c(P', s')\rangle$, with the latter parametrized by the form factors defined in Appendix~\ref{appendix:form factor}. From the previous discussions of the kinematics, it is clear that a scattering process generally occupies a different kinematic region from that of a decay. Thus, to extend the form factors that are commonly convenient for the $\Lambda_c$ weak decays to the scattering processes, their parametrization must be analytic in the proper $q^2$ region, the first crucial criterion mentioned in Sec.~\ref{sec:models}. 

\renewcommand{\theequation}{C.\arabic{equation}}
\section{The amplitude squared of the LFV (LFC) scattering process}
\label{appendix:amplitude}

For the convenience of interested readers and future discussions, we provide here the explicit expression of the amplitude squared $\overline{|\mathcal{M}|}^2 $ of the LFV scattering process $e^-(k)+p(P)\to \mu^-(k')+\Lambda_c(P')$ mediated by the general effective Lagrangian $\mathcal{L}_{\text{eff}}$ (see Eq.~\eqref{eq:Lag_LQ}). To this end, we present it in terms of the kinematic variable $q^2$, the transition form factors (all being functions of $q^2$), the electron beam energy $E$, and the masses of the baryons ($m_p$ and $m_{\Lambda_c}$) and the muon lepton ($m_\mu$). For the LFC scattering process $e^-(k)+p(P)\to e^-(k')+\Lambda_c(P')$, on the other hand, its amplitude can be straightforwardly obtained from above by setting $m_{\mu}$ to zero, since the electron mass can be safely ignored in this case. 
 
With all the operators in Eq.~\eqref{eq:operators} taken into account, the spin-averaged amplitude squared $\overline{|\mathcal{M}|}^2$ of the LFV scattering process $e^-(k)+p(P)\to \mu^-(k')+\Lambda_c(P')$ is given by 
\begin{widetext}
	\begin{align}
		\overline{|\mathcal{M}|}^2&=(|g_{V}^{LL}|^2+|g_{V}^{RR}|^2)\,\overline{|\mathcal{M}|}^2_{V_{LL}-V_{LL}}\!+(|g_{V}^{LR}|^2+|g_{V}^{RL}|^2)\,\overline{|\mathcal{M}|}^2_{V_{LR}-V_{LR}}\!+(|g_{S}^{L}|^2+|g_{S}^{R}|^2)\,\overline{|\mathcal{M}|}^2_{S_{L}-S_{L}}\nonumber\\[0.15cm]
		&+(|g_{T}^{L}|^2+|g_{T}^{R}|^2)\,\overline{|\mathcal{M}|}^2_{T_{L}-T_{L}}+2\text{Re}[g_{V}^{LR}g_{V}^{LL*}+g_{V}^{RL}g_{V}^{RR*}]\,\overline{|\mathcal{M}|}^2_{V_{LR}-V_{LL}}\nonumber\\[0.15cm]
		&+2\text{Re}[g_{V}^{LL}g_{S}^{L*}+g_{V}^{RR}g_{S}^{R*}]\,\overline{|\mathcal{M}|}^2_{V_{LL}-S_{L}}+2\text{Re}[g_{V}^{LR}g_{S}^{L*}+g_{V}^{RL}g_{S}^{R*}]\,\overline{|\mathcal{M}|}^2_{V_{LR}-S_{L}}\nonumber\\[0.15cm]
		&+2\text{Re}[g_{V}^{LL}g_{T}^{L*}+g_{V}^{RR}g_{T}^{R*}]\,\overline{|\mathcal{M}|}^2_{V_{LL}-T_{L}}+2\text{Re}[g_{V}^{LR}g_{T}^{L*}+g_{V}^{RL}g_{T}^{R*}]\,\overline{|\mathcal{M}|}^2_{V_{LR}-T_{L}}\nonumber\\[0.15cm]
		&+2\text{Re}[g_{S}^{L}g_{T}^{L*}+g_{S}^{R}g_{T}^{R*}]\,\overline{|\mathcal{M}|}^2_{S_{L}-T_{L}}\,,
		\label{amplitude}
	\end{align}
\end{widetext}
where the various subscripts attached to the different $\overline{|\mathcal{M}|}^2$ on the right-hand side represent the possible interferences between two operators. For instance, the subscript $V_{LR}$ corresponds to the operator $j_V^LJ^R_V$, while $V_{LL}$ to $j_V^LJ^L_V$. Then, the reduced spin-averaged amplitude squared $\overline{|\mathcal{M}|}^2_{V_{LR}-V_{LL}}$ generated by the interference between the operators $V_{LR}$ and $V_{LL}$ can be written as
\begin{align}
	\overline{|\mathcal{M}|}^2_{V_{LR}-V_{LL}} &\!=\! \frac{1}{4} \sum\limits_{\text{spins}} \langle \mu^-(k') \Lambda_c(P')| j_V^LJ^L_V|e^-(k) p(P) \rangle \,\nonumber \\[0.15cm]
	& \!\times\! \langle \mu^-(k') \Lambda_c(P')| j_V^LJ^R_V|e^-(k) p(P) \rangle^*.
\end{align}
Note that, due to the chiral structures of the lepton and quark currents involved, the reduced amplitudes squared with different subscripts can be identical to each other, \textit{e.g.}, $\overline{|\mathcal{M}|}^2_{V_{LR}-V_{LL}} = \overline{|\mathcal{M}|}^2_{V_{RL}-V_{RR}}$; in this case, only one of them is presented in Eq.~\eqref{amplitude}. The amplitudes associated with other interference terms that are not shown in Eq.~\eqref{amplitude} are all zero, which can be straightforwardly checked according to the chiral structures of the lepton currents involved. For convenience, explicit expressions of the reduced spin-averaged amplitudes squared on the right-hand side of Eq.~\eqref{amplitude} are given, respectively, as 
\begin{widetext}
	\begin{align}
		\overline{|\mathcal{M}|}^2_{V_{LL}-V_{LL}}\!\! &=\!\!\frac{m_\mu^2(m_\mu^2\!-\!q^2)}{8q^4}\Big\{(m_{\Lambda_c}\!\!-\!m_p)^2 \big[(m_{\Lambda_c}\!\!+\!m_p)^2\!-\!q^2\big]f_0^2+(m_{\Lambda_c} \!\!+\! m_p)^2\big[(m_{\Lambda_c}\!\!-\!m_p)^2\!-\!q^2\big]g_0^2\Big\} \nonumber\\[0.15cm]
		&+\left\{\frac{ (m_{\Lambda_c} + m_p)^2}{8[(m_{\Lambda_c} + m_p)^2 - q^2]q^4} f_+^2+\frac{ (m_{\Lambda_c} - m_p)^2}{8[(m_{\Lambda_c} - m_p)^2 - q^2] q^4} g_+^2\right\}\nonumber\\[0.15cm]
		& \times\Big\{4 m_p q^4 (2 E + m_p) (2 E m_p + q^2) + 
		m_\mu^4 (m_p^2 + q^2)^2 + m_{\Lambda_c}^4 m_\mu^2(m_\mu^2 -q^2)\nonumber\\[0.15cm]
		& - 2 m_{\Lambda_c}^2 (m_\mu^2 - q^2) \big[m_\mu^2 (m_p^2 + q^2)-4 E m_p q^2\big] - 
		m_\mu^2 q^2 \big[m_p^4 + 6 m_p^2 q^2 + q^4 \nonumber\\[0.15cm]
		&+ 8 E m_p (m_p^2 + q^2)\big]\Big\}-\frac{1}{4}\left[\frac{f_\perp^2}{(m_{\Lambda_c} + m_p)^2 - q^2}+\frac{g_\perp^2}{(m_{\Lambda_c} - m_p)^2 - q^2}\right]\nonumber\\[0.15cm]
		&\times\Big\{2 m_\mu^4 m_p^2 + m_{\Lambda_c}^4 (q^2-m_\mu^2) 
		+ 2 m_{\Lambda_c}^2 (m_\mu^2 - q^2) (2 E m_p + m_p^2 + q^2) \nonumber\\[0.15cm]
		&- m_\mu^2 (m_p^2 + q^2) (4 E m_p + m_p^2 + q^2) 
		+ q^2 \big[8 E^2 m_p^2 + m_p^4 + q^4 + 4 E m_p (m_p^2 + q^2)\big]\Big\}\nonumber\\[0.15cm]
		&-\frac{m_\mu^2 (m_{\Lambda_c}^2 - m_p^2)}{4q^4}\big[m_{\Lambda_c}^2 (m_\mu^2 - q^2) - m_\mu^2 (m_p^2 + q^2) + 
		q^2 (4 E m_p + m_p^2 + q^2)\big]\nonumber\\[0.15cm]
		&\times(f_0\, f_++g_0\, g_+)\!-\!\frac{1}{2} \big[m_{\Lambda_c}^2 (m_\mu^2 - q^2)\!-\!m_\mu^2 (m_p^2 + q^2) + q^2 (4 E m_p + m_p^2 + q^2)\big]f_\perp\, g_\perp\,, \\[0.2cm]
		\overline{|\mathcal{M}|}^2_{V_{LR}-V_{LR}}\!\! &=\!\!\frac{m_\mu^2(m_\mu^2\!-\!q^2)}{8q^4}\Big\{(m_{\Lambda_c}\!\!-\!m_p)^2 \big[(m_{\Lambda_c}\!\!+\!m_p)^2\!-\!q^2\Big]f_0^2+(m_{\Lambda_c} \!\!+\! m_p)^2\big[(m_{\Lambda_c}\!\!-\!m_p)^2\!-\!q^2\Big]g_0^2\Big\} \nonumber\\[0.15cm]
		&+\Big\{\frac{(m_{\Lambda_c} + m_p)^2}{8\big[(m_{\Lambda_c} + m_p)^2 - q^2\big] q^4} f_+^2+\frac{(m_{\Lambda_c} - m_p)^2}{8\big[(m_{\Lambda_c} - m_p)^2 - q^2\big] q^4} g_+^2\Big\}\nonumber\\[0.15cm]
		& \times\Big\{4 m_p q^4 (2 E + m_p) (2 E m_p + q^2) + 
		m_\mu^4 (m_p^2 + q^2)^2 + m_{\Lambda_c}^4 m_\mu^2(m_\mu^2 -q^2)\nonumber\\
		& - 2 m_{\Lambda_c}^2 (m_\mu^2 - q^2) \big[m_\mu^2 (m_p^2 + q^2)-4 E m_p q^2\big] - 
		m_\mu^2 q^2 \big[m_p^4 + 6 m_p^2 q^2 + q^4 \nonumber\\[0.15cm]
		&+ 8 E m_p (m_p^2 + q^2)\big]\Big\}-\frac{1}{4}\left[\frac{f_\perp^2}{(m_{\Lambda_c} + m_p)^2 - q^2}+\frac{g_\perp^2}{(m_{\Lambda_c} - m_p)^2 - q^2}\right]\nonumber\\[0.15cm]
		&\times\Big\{2 m_\mu^4 m_p^2 + m_{\Lambda_c}^4 ( q^2-m_\mu^2) 
		+ 2 m_{\Lambda_c}^2 (m_\mu^2 - q^2) (2 E m_p + m_p^2 + q^2) \nonumber\\[0.15cm]
		&- m_\mu^2 (m_p^2 + q^2) (4 E m_p + m_p^2 + q^2) 
		+ q^2\big[8 E^2 m_p^2 + m_p^4 + q^4 + 4 E m_p (m_p^2 + q^2)\big]\Big\}\nonumber\\[0.15cm]
		&-\frac{ m_\mu^2 (m_{\Lambda_c}^2 - m_p^2)}{4q^4}\big[m_{\Lambda_c}^2 (m_\mu^2 - q^2) - m_\mu^2 (m_p^2 + q^2) + 
		q^2 (4 E m_p + m_p^2 + q^2)\big]\nonumber\\[0.15cm]
		&\times(f_0\, f_++g_0\, g_+)\!+\!\frac{1}{2} \big[m_{\Lambda_c}^2 (m_\mu^2 - q^2)\!-\!m_\mu^2 (m_p^2 + q^2) + q^2 (4 E m_p + m_p^2 + q^2)\big]f_\perp\, g_\perp\,, \\[0.2cm]
		\overline{|\mathcal{M}|}^2_{S_{L}-S_{L}}\!&=\!\frac{(m_\mu^2\!-\!q^2)}{8m_{c}^2}\Big\{(m_{\Lambda_c}\!- \!m_p)^2  \big[(m_{\Lambda_c}\!+\!m_p)^2\!-\!q^2\big]f_0^2+(m_{\Lambda_c}\!+\!m_p)^2  \big[(m_{\Lambda_c}\!-\!m_p)^2- q^2\big]g_0^2\Big\}\,,\\[0.2cm]
		\overline{|\mathcal{M}|}^2_{T_{L}-T_{L}}&=-\left[\dfrac{2h_+^2}{(m_{\Lambda_c}+m_p)^2-q^2}+\dfrac{2\tilde{h}_+^2}{(m_{\Lambda_c}-m_p)^2-q^2}\right]\Big\{4 m_\mu^4 m_p^2 + m_{\Lambda_c}^4 (-m_\mu^2 + q^2) \nonumber\\[0.15cm]
		&+ 2 m_{\Lambda_c}^2 (m_\mu^2 - q^2) (4 E m_p + m_p^2 + q^2) + q^2 (4 E m_p + m_p^2 + q^2)^2\nonumber\\[0.15cm]
		& - m_\mu^2 \big[m_p^4 + 6 m_p^2 q^2 + q^4 + 8 E m_p (m_p^2 + q^2)\big]\Big\}\nonumber\\[0.15cm]
		&+\left\{\dfrac{4h_\perp^2(m_{\Lambda_c}+m_p)^2}{\big[(m_{\Lambda_c}+m_p)^2-q^2\big]q^4}+\dfrac{4\tilde{h}_\perp^2(m_{\Lambda_c}-m_p)^2}{\big[(m_{\Lambda_c}-m_p)^2-q^2\big]q^4}\right\}
		\Big\{2 m_p (2 E + m_p) q^4 (2 E m_p + q^2)\nonumber\\[0.15cm]
		& - m_\mu^2 q^2 (m_p^2 + q^2) (4 E m_p + m_p^2 + q^2) + m_{\Lambda_c}^4 m_\mu^2 (m_\mu^2 - q^2) \nonumber\\[0.15cm]
		&+m_\mu^4 (m_p^4 + q^4) - 2 m_{\Lambda_c}^2 (m_\mu^2 - q^2) \big[m_\mu^2 (m_p^2 + q^2)-2 E m_p q^2\big]\Big\}\nonumber\\[0.15cm]
		&-\frac{8 m_\mu^2 (m_{\Lambda_c}^2 - m_p^2)}{q^4} \big[m_{\Lambda_c}^2 (m_\mu^2 - q^2) - m_\mu^2 (m_p^2 + q^2) + 
		q^2 (4 E m_p + m_p^2 + q^2)\big]h_\perp\,\tilde{h}_\perp\,, \\[0.2cm]
		\overline{|\mathcal{M}|}^2_{V_{LR}-V_{LL}}\!\! &=\!\!\frac{m_\mu^2(m_\mu^2\!-\!q^2)}{8q^4}\Big\{(m_{\Lambda_c}\!\!-\!m_p)^2 \big[(m_{\Lambda_c}\!\!+\!m_p)^2\!-\!q^2\big]f_0^2\!-\!(m_{\Lambda_c} \!\!+\! m_p)^2\big[(m_{\Lambda_c}\!\!-\!m_p)^2\!-\!q^2\big]g_0^2\Big\} \nonumber\\[0.15cm]
		&+\left\{\frac{(m_{\Lambda_c} + m_p)^2}{8\big[(m_{\Lambda_c} + m_p)^2 - q^2\big] q^4} f_+^2-\frac{(m_{\Lambda_c} - m_p)^2}{8\big[(m_{\Lambda_c} - m_p)^2 - q^2\big]q^4} g_+^2\right\}\nonumber\\[0.15cm]
		& \times\Big\{4 m_p q^4 (2 E + m_p) (2 E m_p + q^2) + 
		m_\mu^4 (m_p^2 + q^2)^2 + m_{\Lambda_c}^4 m_\mu^2(m_\mu^2 -q^2)\nonumber\\[0.15cm]
		& - 2 m_{\Lambda_c}^2 (m_\mu^2 - q^2) \big[m_\mu^2 (m_p^2 + q^2)-4 E m_p q^2\big] - 
		m_\mu^2 q^2 \big[m_p^4 + 6 m_p^2 q^2 + q^4 \nonumber\\[0.15cm]
		&+ 8 E m_p (m_p^2 + q^2)\big]\Big\}-\frac{1}{4}\left[\frac{f_\perp^2}{(m_{\Lambda_c} + m_p)^2 - q^2}-\frac{g_\perp^2}{(m_{\Lambda_c} - m_p)^2 - q^2}\right]\nonumber\\[0.15cm]
		&\times\Big\{2 m_\mu^4 m_p^2 + m_{\Lambda_c}^4 (q^2-m_\mu^2) 
		+ 2 m_{\Lambda_c}^2 (m_\mu^2 - q^2) (2 E m_p + m_p^2 + q^2) \nonumber\\[0.15cm]
		&- m_\mu^2 (m_p^2 + q^2) (4 E m_p + m_p^2 + q^2) 
		+ q^2 \big[8 E^2 m_p^2 + m_p^4 + q^4 + 4 E m_p (m_p^2 + q^2)\big]\Big\}\nonumber\\[0.15cm]
		&-\frac{m_\mu^2 (m_{\Lambda_c}^2 - m_p^2)}{4q^4}\big[m_{\Lambda_c}^2 (m_\mu^2 - q^2) - m_\mu^2 (m_p^2 + q^2) + 
		q^2 (4 E m_p + m_p^2 + q^2)\big](f_0\, f_+-g_0\, g_+)\,,\\[0.2cm]
		\overline{|\mathcal{M}|}^2_{V_{LL}-S_{L}}&=\frac{m_\mu}{8m_c\,q^2}
		\Big\{(m_{\Lambda_c}^2-m_p^2)
		\big[m_{\Lambda_c}^2 (m_\mu^2 - q^2) - m_\mu^2 (m_p^2 + q^2) + q^2 (4 E m_p + m_p^2 + q^2)\big]\nonumber\\[0.15cm]
		&\times(f_0\,f_+ + g_0\,g_+)-(m_\mu^2 - q^2)\big[(m_{\Lambda_c} - m_p)^2((m_{\Lambda_c} + m_p)^2 - q^2)f_0^2\nonumber\\[0.15cm]
		&+(m_{\Lambda_c}+ m_p)^2((m_{\Lambda_c} - m_p)^2 - q^2)g_0^2\big]\Big\}\,, \\[0.2cm]
		\overline{|\mathcal{M}|}^2_{V_{LR}-S_{L}}&=\frac{ m_\mu}{8m_c\,q^2}
		\Big\{(m_{\Lambda_c}^2-m_p^2)
		\big[m_{\Lambda_c}^2 (m_\mu^2 - q^2) - m_\mu^2 (m_p^2 + q^2) + q^2 (4 E m_p + m_p^2 + q^2)\big]\nonumber\\[0.15cm]
		&\times(f_0\,f_+ - g_0\,g_+)-(m_\mu^2 - q^2)\big[(m_{\Lambda_c} - m_p)^2((m_{\Lambda_c} + m_p)^2 - q^2)f_0^2\nonumber\\[0.15cm]
		&-(m_{\Lambda_c}+ m_p)^2((m_{\Lambda_c} - m_p)^2 - q^2)g_0^2\big]\Big\}\,,\\[0.2cm]
		\overline{|\mathcal{M}|}^2_{V_{LL}-T_{L}}&=-\frac{m_\mu}{2q^2}\Big\{\big[m_{\Lambda_c}^2 (m_\mu^2 - q^2) - m_\mu^2 (m_p^2 + q^2) + q^2 (4 E m_p + m_p^2 + q^2)\big]\nonumber\\[0.15cm]
		&\times\big[(m_{\Lambda_c} - m_p) (f_0\,h_+ +2f_\perp\,\tilde{h}_\perp)
		+(m_{\Lambda_c} + m_p) (g_0 \,\tilde{h}_+ + 2g_\perp\,h_\perp)\big]\nonumber\\[0.15cm]
		&-(m_\mu^2 - q^2)\big[(m_{\Lambda_c} + m_p)((m_{\Lambda_c} - m_p)^2 - q^2)(f_+\, h_+ + 2f_\perp\, h_\perp) \nonumber\\[0.15cm]
		&+(m_{\Lambda_c}- m_p)((m_{\Lambda_c}+ m_p)^2 - q^2)(g_+\,\tilde{h}_+ + 2g_\perp\, \tilde{h}_\perp)\big]\Big\}\,, \\[0.2cm]
		\overline{|\mathcal{M}|}^2_{V_{LR}-T_{L}}&=-\frac{m_\mu}{2q^2}\Big\{\big[m_{\Lambda_c}^2 (m_\mu^2 - q^2) - m_\mu^2 (m_p^2 + q^2) + q^2 (4 E m_p + m_p^2 + q^2)\big]\nonumber\\[0.15cm]
		&\times\big[(m_{\Lambda_c} - m_p) (f_0\,h_+ +2f_\perp\,\tilde{h}_\perp)
		-(m_{\Lambda_c} + m_p) (g_0 \,\tilde{h}_+ + 2g_\perp\,h_\perp)\big]\nonumber\\[0.15cm]
		&-(m_\mu^2 - q^2)\big[(m_{\Lambda_c} + m_p)((m_{\Lambda_c} - m_p)^2 - q^2)(f_+\, h_+ + 2f_\perp\, h_\perp) \nonumber\\[0.15cm]
		&-(m_{\Lambda_c}- m_p)((m_{\Lambda_c}+ m_p)^2 - q^2)(g_+\,\tilde{h}_+ + 2g_\perp\, \tilde{h}_\perp)\big]\Big\}\,, \\[0.2cm]
		\overline{|\mathcal{M}|}^2_{S_{L}-T_{L}}&=\frac{1}{2m_{c}} \Big\{m_{\Lambda_c}^2 (m_\mu^2 - q^2) - m_\mu^2 (m_p^2 + q^2) + 
		q^2 (4 E m_p + m_p^2 + q^2)\nonumber\\[0.15cm]
		&\times \big[(m_{\Lambda_c} - m_p) f_0\, h_+ + (m_{\Lambda_c} + m_p) g_0 \tilde{h}_+\big]\Big\}\,.
	\end{align}
\end{widetext}

\bibliographystyle{apsrev4-1}
\bibliography{reference}

\end{document}